\documentclass[twocolumn]{aa} % for a paper on 1 column  
\usepackage{xcolor}
\usepackage{graphicx}
\usepackage[normalem]{ulem}
\usepackage{txfonts}
\usepackage{hyperref}
\hypersetup{colorlinks=true, linkcolor={red}}
\begin{document}

\title{Adiabatic-radiative shock systems in YSO jets and novae outflows}
%   \subtitle{I. Overviewing the $\kappa$-mechanism}

\titlerunning{Shocks in YSO \& novae}
\authorrunning{del Valle, Araudo \& Suzuki-Vidal}

   \author{M.V. del Valle
          \inst{1},
          A. Araudo\inst{2,3,4}
          \and
          F. Suzuki-Vidal\inst{5}
          %\fnmsep\thanks{Just to show the usage
          %of the elements in the author field}
          }

   \institute{{Universidade de São Paulo, Instituto de Astronomia, Geofísica e
Ciências Atmosféricas, Brazil}\\
              \email{mvdelvalle@usp.br}
         \and
         Laboratoire Univers et Particules de Montpellier (LUPM) Universit{\'e} Montpellier, CNRS, France
         \and
             ELI Beamlines, Institute of Physics, Czech Academy of Sciences, 
25241 Doln{\'\i} B\v re\v zany, Czech Republic
\and
Astronomical Institute, Czech Academy of Sciences,
Bo\v{c}n\'{\i} II 1401, CZ-141\,00 Prague, Czech Republic\\
             \email{anabella.araudo@eli-beams.eu}
             \and
Blackett Laboratory, Imperial College London, London SW7 2BW, United Kingdom\\
             \email{f.suzuki@imperial.ac.uk}
%             \thanks{The university of heaven temporarily does not
%                     accept e-mails}
             }

   \date{Received ; accepted }

% \abstract{}{}{}{}{} 
% 5 {} token are mandatory
 
  \abstract
   {The termination regions of non-relativistic jets in protostars and supersonic outflows in classical novae are nonthermal emitters. This has been confirmed by radio and gamma-ray detection, respectively. A two-shock system is expected to be formed in the termination region where the jet/outflow material and the ambient medium impact. Given the high densities in these systems, radiative shocks are expected to form. However, in the presence of high velocities, the formation of adiabatic shocks is also possible. A case of interest is when the two types of shocks occur simultaneously. Adiabatic shocks are more efficient at particle acceleration while radiative shocks strongly compress the gas. Furthermore, a combined adiabatic-radiative shock system is very prone to develop instabilities in the contact discontinuity leading to mixing, turbulence and density enhancement. Additionally, these dense non-relativistic jets/outflows are excellent candidates for laboratory experiments as demonstrated by magnetohydrodynamics scaling.}
  % aims heading  
   {We aim at studying the combination of adiabatic and radiative shocks in protostellar jets and novae outflows. We focus on determining the conditions under which this combination is feasible together with its physical implications.}
   {We perform an analytical study of the shocks in both types of sources for a set of parameters by comparing cooling times and propagation velocities. We also estimate the timescales for the growth of instabilities in the contact discontinuity separating both shocks. The hydrodynamical evolution of a jet colliding with an ambient medium is studied with 2D numerical simulations confirming our initial theoretical estimates.}
   {We show that for a wide set of observationally constrained parameters the combination of an adiabatic and a radiative shock is possible at the working surface of the termination region in jets from young stars and novae outflows. We find that instabilities are developed at the contact discontinuity, mixing the shocked materials. Additionally, we explore the magnetohydrodynamic parameter scaling required for studying protostellar jets and novae outflows using laboratory experiments on laser facilities.}
   {The coexistence of an adiabatic and a radiative shock is expected at the termination region of protostellar jets and novae outflows. This scenario is very promising for particle acceleration and gamma-ray emission. The parameters for scaled laboratory experiments are very much in line with plasma conditions achievable in currently operating high-power laser facilities. This opens the door to new means for studying novae outflows never considered before.}

   \keywords{shock waves --
               ISM: jets and outflows --
               stars: jets --
               novae, cataclysmic variables --
               instabilities --
               radiation mechanisms: nonthermal
               }

    \maketitle
%
%-------------------------------------------------------------------

\section{Introduction}
Astrophysical jets and outflows from pc to Mpc scales are very common in the Universe \cite[e.g.][]{LIVIO1999, DEGOUVEIADALPINO2005908}. Shocks in non-relativistic jets and outflows such as those in protostars and novae are expected to be radiative given their large densities. Radiative shocks are not efficient particle accelerators, however, there is evidence of nonthermal emission. Synchrotron radio emission has been detected both in protostellar jets \citep[e.g.,][]{Carlos_Sci,2019ApJ...885L...7F} and novae \citep[e.g.,][]{Vlasov_Metzger_2016}, whereas gamma-rays have only been detected in the latter \citep{Fermi_Novae}.  Accelerating synchrotron radio emitting electrons is possible for almost any kind of shock. However, accelerating TeV particles for gamma-ray emission is not trivial for radiative shocks. 
\cite{2018MNRAS.479..687S} showed that a working surface composed of two radiative shocks can accelerate ions with an efficiency of $\sim 0.01$ when non-linear thin shell instabilities take place at the contact discontinuity. In contrast, non-relativistic adiabatic shocks have an efficiency 10 times greater \citep{Caprioli_14a}.

We are interested in the combination of adiabatic and radiative shocks in the termination region of non-relativistic jets/outflows. The self-similar  dynamics of this configuration of shocks was studied by the first time by \cite{Antoine-21}. The advantage of this configuration of shocks is that, whereas particles are accelerated in the adiabatic shock, the downstream region of the radiative shock acts as a dense target for gamma-ray and neutrino emission by inelastic proton-proton (pp) collisions and via relativistic Bremsstrahlung.  In addition, the radiation field from the downstream region of the radiative shock can ionize the jet/outflow plasma, increasing the efficiency of particle acceleration from the adiabatic shock. This termination region is prone to the growth of hydrodynamic instabilities. It is now possible to study some of these instabilities through high-energy density laboratory experiments, for instance Rayleigh-Taylor \citep{Kuranz2018_Nature_RT}, Kevin-Helmholtz \citep{Doss2015_KH} and thermal instabilities \citep{Suzuki-Vidal_15}.

In this work we focus on protostellar jets and classical novae outflows. For simplicity, we will refer to both outflows as \textit{jets}. In both cases the initial jet flow moves with velocities $v_{\rm j}\sim 100-1000$~km~s$^{-1}$, but in ambient media with different densities (see Table~\ref{Tab:0}). We show that for certain combinations of jet velocities and ambient ($n_{\rm a}$) and jet ($n_{\rm j}$) densities, the working surface in the termination region is composed of an adiabatic and a radiative shock. We also show that this combination leads to a fast growth of hydrodynamical instabilities and therefore a significant level of mixing. This situation is very promising for gamma-ray emission through pp inelastic collisions and relativistic Bremsstrahlung. 

The paper is organized as follows: in Sec.~\ref{jetshocks} we describe the properties of the shocks in the termination region. In Sec.~\ref{instabilities} we perform a stability study of the leading working surface. In Sec.~\ref{numerical} we perform numerical simulations with the freely distributed code PLUTO.  In Sec.~\ref{discussion} we discuss the results and implications for the expected gamma-ray emission. In Sec.~\ref{scaling} we discuss magnetohydrodyanic (MHD) scaling of YSO jets and novae outflows to laboratory experiments. Finally, in Sec.~\ref{conclusions} we present the summary and conclusions of this work.

\begin{table*}
\caption{Typical flow and ambient parameters for YSO jets and novae outflows.}
\begin{tabular}{lll}
\hline
Parameter & YSO & Novae \\
 \hline
Velocity $[v_{\rm j}]=\rm km\,s^{-1}$ & $100-1000$ & 1000\\
Mass loss rate $[\dot{M}]$ & $10^{-8}-10^{-5}$ [$M_{\odot}\,\rm yr^{-1}$] & $10^{-6}-10^{-3}$ [$M_{\odot}\,\rm wk^{-1}$]\\
Radius $[R_j]=\rm$ cm & $10^{16}$ & $6\times10^{13}$\\
Density  $[n_{\rm j}]=\rm cm^{-3}$ & $1-10^3$ & $10^{9}$\\
Ambient density  $[n_{\rm a}]=\rm cm^{-3}$ & 10$ - 10^5$ & $10^{11}$\\
\hline
\label{Tab:0}
\end{tabular}
\end{table*}

\section{Shocks in the termination region}\label{jetshocks}

\begin{figure}[b]
\centering
\includegraphics[width=0.52\textwidth]{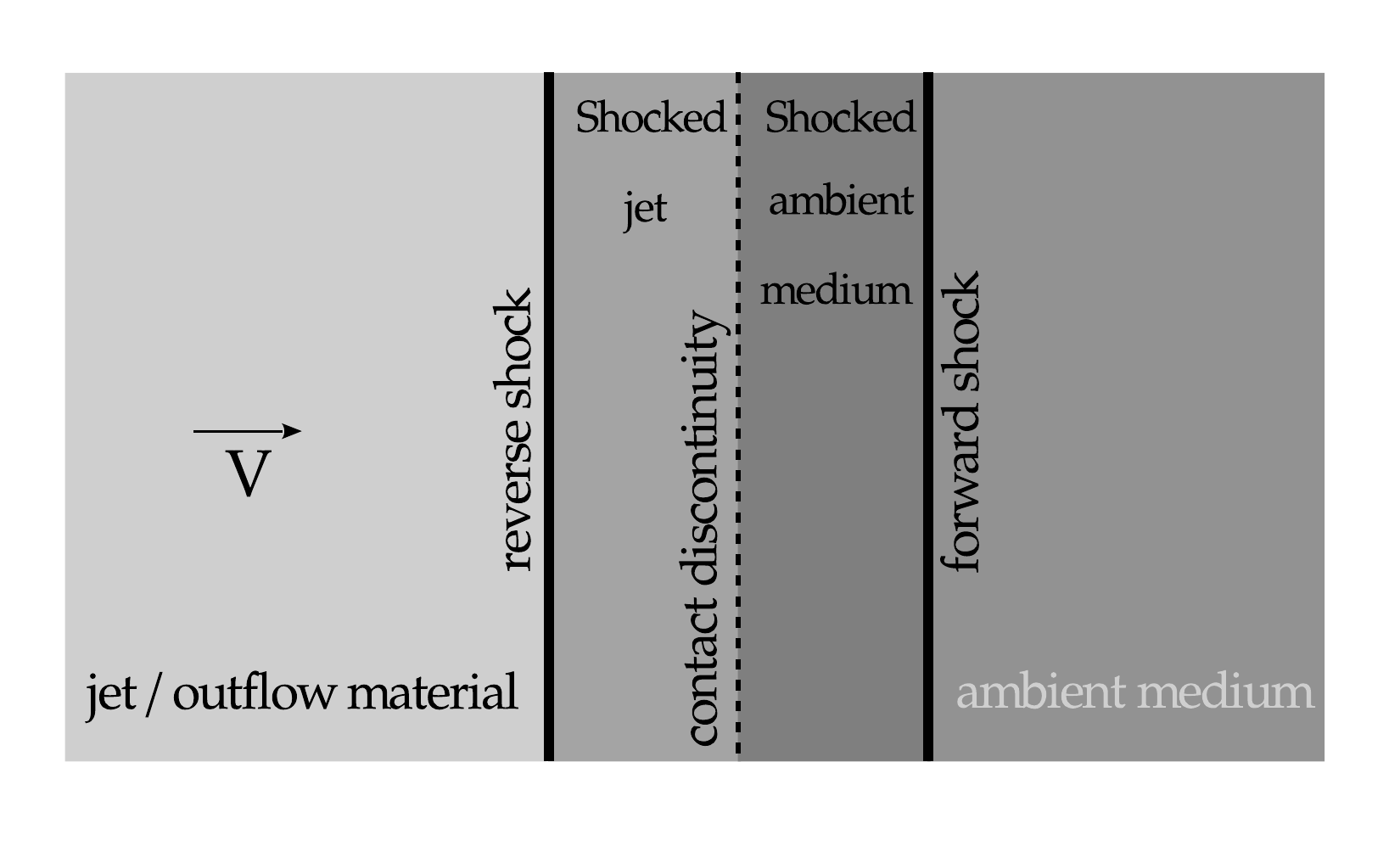}
\caption{Schematic diagram showing the formation of forward and reverse shocks in the termination region of YSO jets and novae outflows interacting with an ambient medium. {The arrow indicates the fluid velocity direction in the laboratory frame.}}
\label{fig0}
\end{figure}

Protostellar jets and novae outflows are supersonic and therefore their termination region is made of a bow shock (or forward shock) in the ambient medium, and a Mach disc (or reverse shock) moving into the initial jet flow (Fig.~\ref{fig0}). In the uni-dimensional flow approximation, the bow shock moves into the ambient medium at $v_{\rm bs} \sim v_{\rm j}/(1 + 1/\sqrt{\chi})$, where $\chi \equiv n_{\rm j}/n_{\rm a}$ is the jet to ambient density contrast \cite[e.g.][]{Alex_Jorge_WS, Hartigan_MachDisc}. The reverse shock in the jet moves at $v_{\rm rs} = v_{\rm j} - 3 v_{\rm bs}/4$. In ``heavy'' jets ($\chi > 1$), the bow shock is faster than the Mach disc, whereas in ``light'' jets ($\chi < 1$), the reverse shock is faster than the bow shock. In particular, 
$v_{\rm rs}\sim v_{\rm j}$ and $v_{\rm bs}\sim v_{\rm j}\sqrt{\chi}$ when $\chi \ll 1$, whereas $v_{\rm bs}\sim v_{\rm j}$ when $\chi\gg1$.

The jet density in the termination region can be calculated from conservation of mass as \citep{Rodriguez_Kamenetzky_2017}
\begin{equation}\label{dens_z}
\frac{n_{\rm j}}{\rm cm^{-3}} \approx 150
\left(\frac{\dot M_{\rm i}}{10^{-6}\,\rm M_{\odot}\,yr^{-1}}\right)
\left(\frac{v_{\rm j}}{1000~\rm km\,s^{-1}}\right)^{-1}
\left(\frac{R_{\rm j}}{10^{16}\,\rm cm}\right)^{-2},
\end{equation}

where $\dot M_i$ is the ionized mass loss rate and $R_{\rm j}$ is the radius of the section of the jet in the termination region.  

In the strong shock approximation, the  plasma is compressed by a factor of 4 and the temperature is $T_{\rm ps} \sim 2\times10^5(v_{\rm sh}/100\,{\rm km\,s^{-1}})^2$~K immediately after the shock front. Establishing the nature of the shocks, weather they are adiabatic or radiative, can be done by comparing the advection (escape) time-scale $t_{\rm esc} \sim R_{\rm j}/(v_{\rm sh}/4)$  to the cooling time-scale

\begin{equation}\label{tcool}
t_{\rm cool} = \frac{(3/2) k_{\rm B}T}{n_{\rm a}\, \Lambda(T)}
\end{equation}
where $\Lambda(T)$ is the cooling function which depends strongly on the temperature $T$. For the case of novae, characterised by high shock velocities and thus high temperatures, we use free-free emission which is given by $\Lambda(T_{\rm ps})=2\times10^{-27}T_{\rm ps}^{1/2}$ \,erg\,cm$^{3}$\,s$^{-1}$. Shocks in YSO jets have lower velocities and thus metal-line cooling dominates \citep[e.g.,][]{1993ApJS...88..253S}.
 
Equivalently, we can compare the (thermal) cooling length $l_{\rm cool}=  t_{\rm cool}v_{\rm sh}/4$ and $R_{\rm j}$. The condition $l_{\rm cool} > R_{\rm j}$ \citep{Heathcote_1998} 
for a YSO shock  to be adiabatic  can be rewritten as $v_{\rm sh} > v_{\rm sh, ad}$, where 

\begin{equation}\label{cond_th0}
    \frac{v_{\rm sh, ad}}{\rm km~s^{-1}} \simeq 650
    \left(\frac{n_{\rm a}}{10^4\,\rm cm^{-3}}\right)^\frac{2}{9}
    \left(\frac{R_{\rm j}}{10^{16}\,\rm cm}\right)^\frac{2}{9}.
\end{equation}
 
In order to study the nature of the two shocks in the termination region, i.e.~whether they are adiabatic or radiative, we sample the parameter space for $n_{\rm j}$ and $n_{\rm a}$ for the cases of YSO jets and novae outflows presented in Table \ref{Tab:0}. 

\subsection{Young stellar objects}

Stars are formed within dense molecular clouds, accreting matter onto the central protostar with the formation of a circumstellar disc and bipolar jets. These ejections are collimated flows of disc/stellar matter accelerated by magnetic field lines \citep{Blandford_Payne, Shu}, and moving with speeds $v_{\rm j}\sim 100-1000$~km~s$^{-1}$  into the ambient molecular cloud. Molecular matter from the cloud is entrained by the jet, forming molecular outflows. 

\begin{figure*}%[h!]
\centering
\includegraphics[width=0.3\textwidth,trim=0cm 0cm 0cm 0cm, clip=true,angle=270]{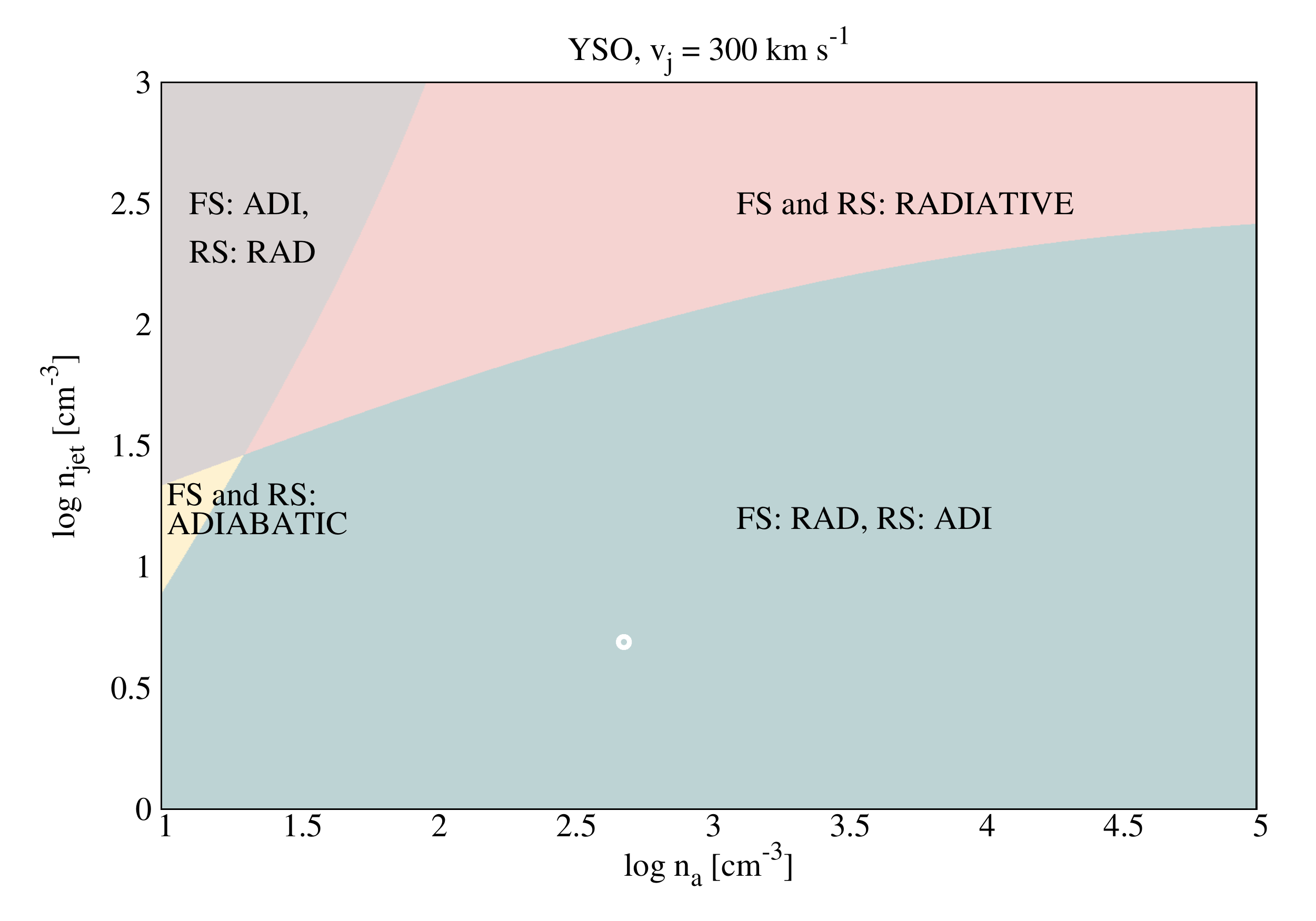}
\includegraphics[width=0.3\textwidth,trim=0cm 0cm 0cm 0cm, clip=true,angle=270]{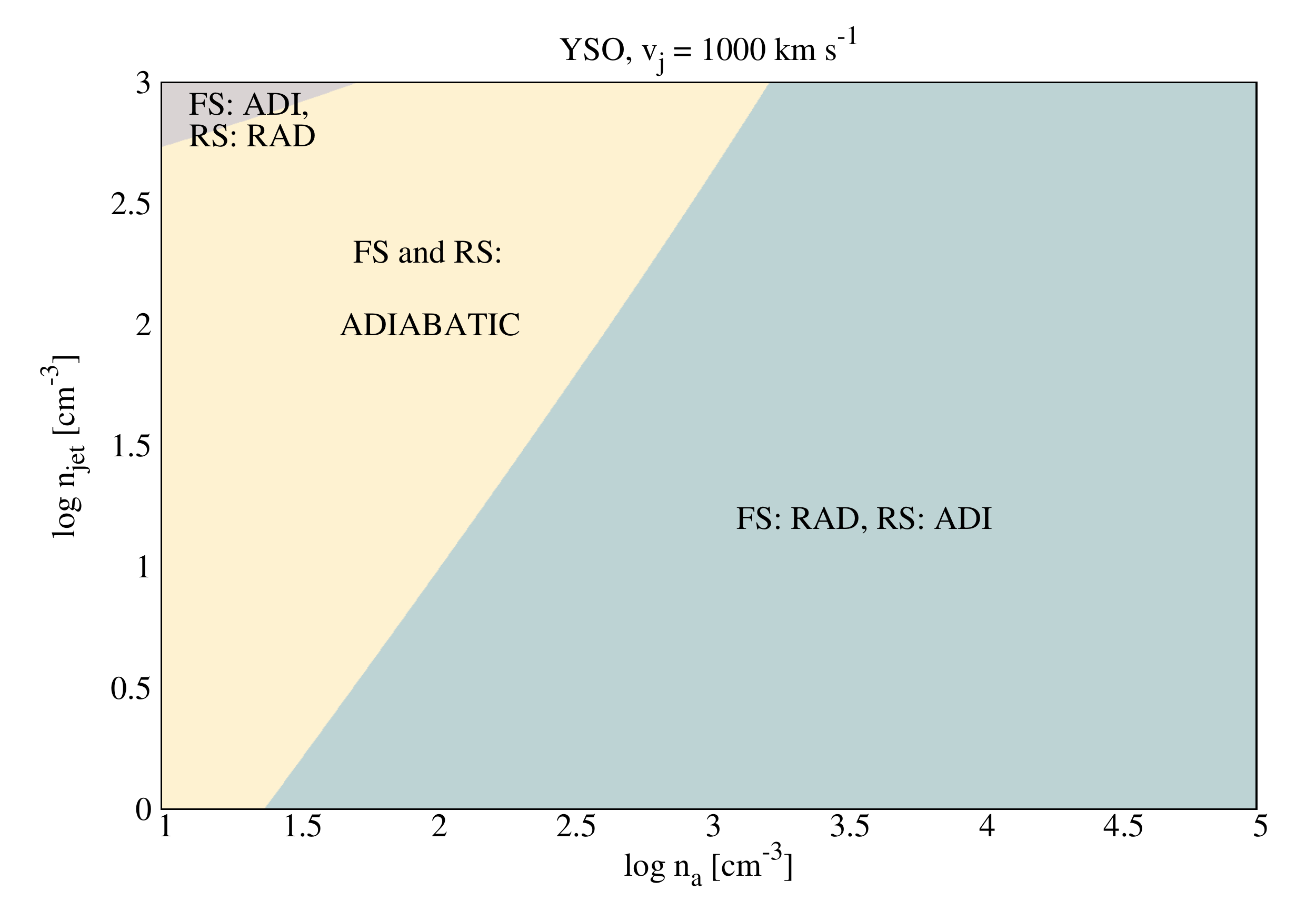}
\caption{Shock nature diagnostic in the termination region of a YSO jet for different jet and ambient densities, for $v_{\rm j} = 300$ (left) and $1000$\,km\,s$^{-1}$ (right). The pink region indicates when the forward shock (FS) and the reverse shock (RS) are radiative while the larger green region indicates when the forward shock is radiative and the reverse shock is adiabatic, the gray color indicates an adiabatic forward shock and a radiative reverse shock while the yellow area shows when both shocks are adiabatic. The white dot indicates the parameters used in the numerical simulations in Section\,\ref{numerical}.}
\label{fig1}
\end{figure*}

The ionized mass loss rate in YSO jets is $10^{-8}\le \dot M_i\le 10^{-5}$~M$_{\odot}$~yr$^{-1}$, and the width of the jet in the termination region is characteristically $R_{\rm j} \sim 10^{16}$~cm. By inserting these values in Eq.~(\ref{dens_z}) we find $n_{\rm j}$ in the range $1-10^3$\,cm$^{-3}$.  The ambient medium is the molecular cloud where the protostar is embedded, and typical values of $n_{\rm a}$ are in the range $10-10^5$\,cm$^{-3}$. This results in values $10^{-5} \le \chi \le 10$. We consider $v_{\rm j}=300$ and $1000$\,km\,s$^{-1}$ and use the condition given by Eq.~(\ref{cond_th0}) to classify the shocks. The results are shown in Figure~\ref{fig1}.
 
The green region in Figure~\ref{fig1} indicates when the bow shock is radiative and the reverse shock is adiabatic, which is the combination of interest of this work. The left plot corresponds to  $v_{\rm j} = 300$\,km\,s$^{-1}$ and the right one to a faster jet  with $v_{\rm j} = 1000$\,km\,s$^{-1}$. The pink region indicates when both shocks cool efficiently, i.e. both shocks are radiative. The gray region indicates an adiabatic forward shock and a radiative reverse shock, while the yellow region shows when both shocks are adiabatic. The particular case for $n_{\rm a} = 500$\,cm$^{-3}$ and $n_{\rm j} = 5$\,cm$^{-3}$  for $v_{\rm j} = 300$\,km\,s$^{-1}$ ,marked with a white dot in the figure, is investigated in detail with 2D numerical simulations in Section\,\ref{numerical}. For faster YSO jets, $v_{\rm j} = 1000$\,km\,s$^{-1}$, the green region is larger, and in the range of densities studied here there is no region where both shocks are radiative (no pink region). We can conclude that for large a range of parameters for the ambient medium and the jet, we can expect the formation of adiabatic-radiative shocks that are of particular interest in this work.

\begin{figure*}%[h!]
\centering
\includegraphics[width=0.33\textwidth,trim=1cm 0cm 1cm 2cm, clip=true,angle=270]{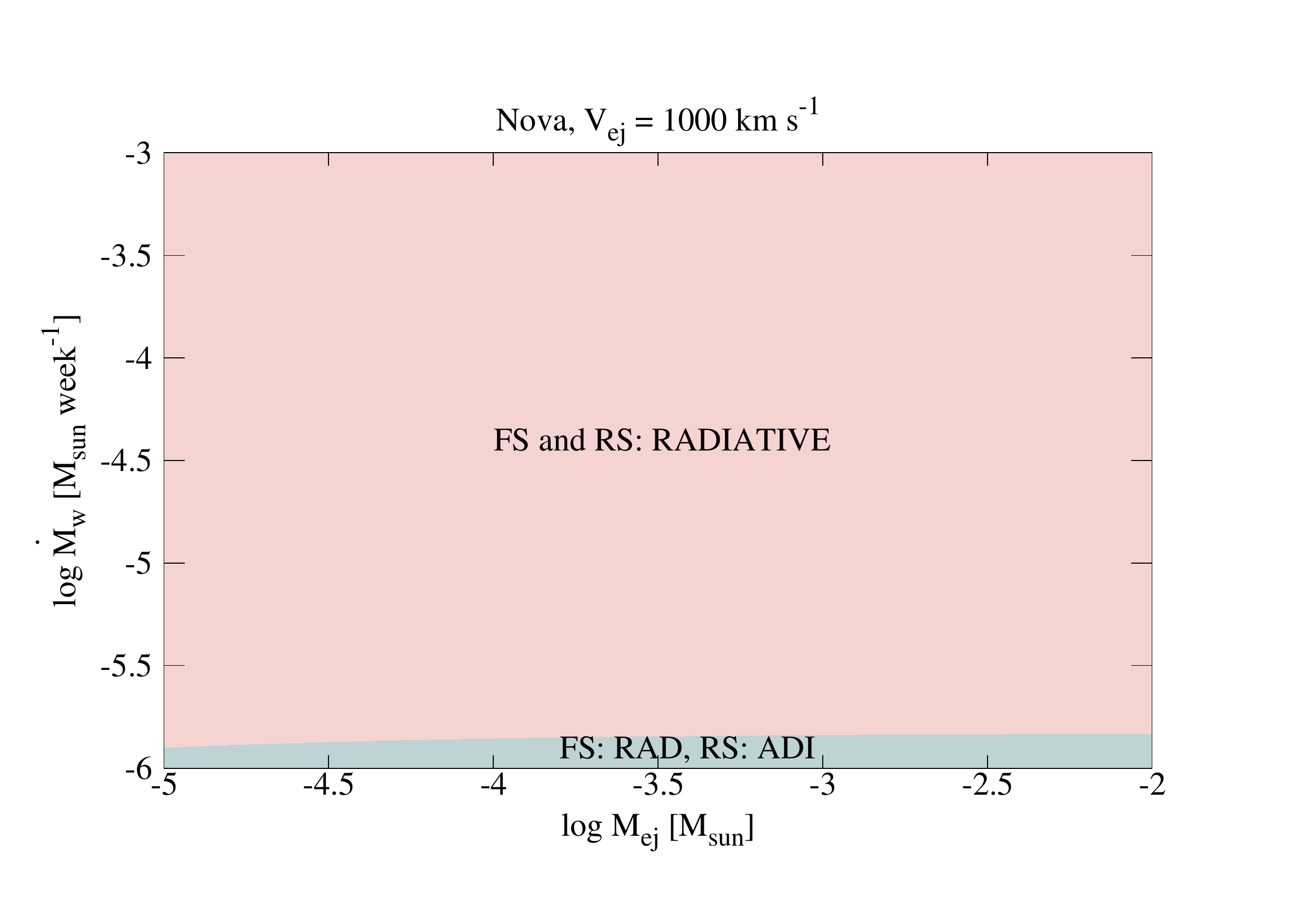}
\includegraphics[width=0.33\textwidth,trim=1cm 0cm 1cm 2cm, clip=true,angle=270]{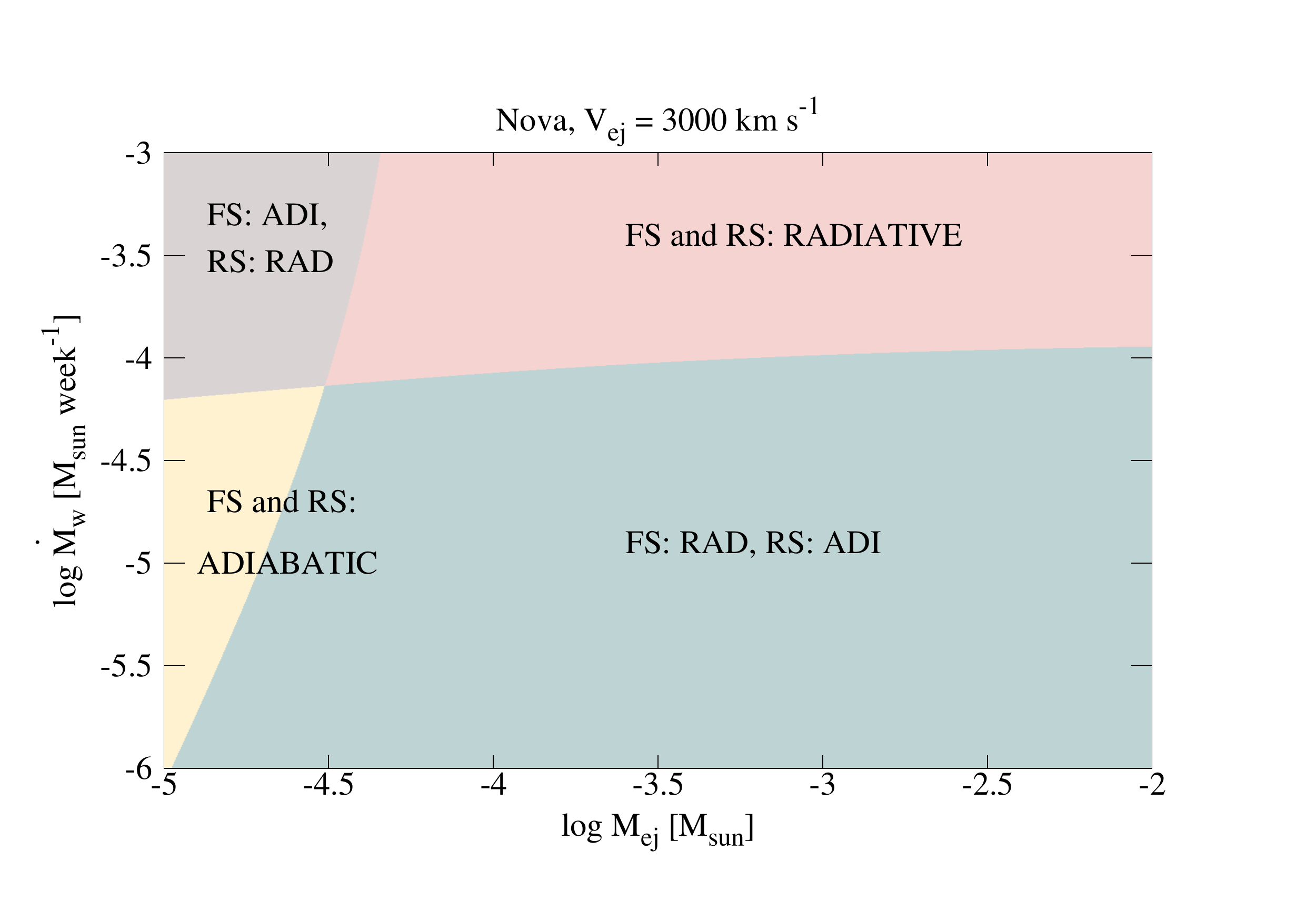}
\caption{Shock nature diagnostic in a nova forward-reverse system, with $V_{\rm ej} = 1000$ %\,km\,s$^{-1}$ 
({\it left}) and  $3000$\,km\,s$^{-1}$ ({\it right}). The color coding is the same as in Figure\,\ref{fig1}.  
}
\label{fig2}
\end{figure*}

\subsection{Classical novae}\label{nova}

In order to study the shocks produced in classic novae we follow the model of \cite{2015MNRAS.450.2739M}; see also \citet{2016MNRAS.457.1786M}. The shocks in novae outflows are formed from the collision of a slow shell ejection with velocity $V_{\rm ej} \sim 1000\,$km\,s$^{-1}$, produced in the thermonuclear runaway, and a faster outflow or continuous wind {of velocity $V_{\rm w}$ $\sim$ 2$V_{\rm ej}$} that follows the ejecta within a few days. The collision, as in the case of YSO jets, produces a system of two shocks: forward and reverse.

The forward shock propagates through the slow shell and the reverse shock moves back through the wind. The shock velocities depend on the density ratio of the colliding media. The expressions are the same as the ones presented at the beginning of this section with $v_{\rm j} = V_{\rm w} - V_{\rm ej}$. The slow shell density is $n_{\rm ej} \approx M_{\rm ej}/(4\pi R_{\rm ej}^{3}f_{\Delta\Omega}m_{p})$, where $M_{\rm ej}$ is the mass of the ejecta, $R_{\rm ej} = V_{\rm ej}t_{\rm ej}$ its expansion radius, and $f_{\Delta\Omega} \sim 0.5$ is a filling factor related to the geometry of the model, i.e.~the fraction of the total solid-angle subtended by the outflow. By considering typical values for the slow ejecta we obtain 
\begin{equation}\label{ej}
\frac{n_{\rm ej}}{\rm cm^{-3}} \approx 10^{10} 
\left(\frac{f_{\Delta\Omega}}{0.5}\right)^{-1}
\left(\frac{M_{\rm ej}}{10^{-4}M_{\odot}}\right)
\left(\frac{V_{\rm ej}}{1000\,\rm km\,s^{-1}}\right)^{-3}
\left(\frac{t_{\rm ej}}{2\rm wk}\right)^{-3}.
\end{equation}
The typical mass-loss rate of the fast wind is $\dot M_{\rm w} = 10^{-5}$\,M$_{\odot}$\,wk$^{-1}$, giving a wind density for an outflow with radius $R_{\rm ej}$

\begin{equation}\label{w}
  \frac{n_{\rm w}}{\rm cm^{-3}} = 2\times10^9\left(\frac{\dot M_{\rm w}}{10^{-5}M_{\odot}\,\rm wk^{-1}}\right)
  \left(\frac{V_{\rm ej}}{1000\,\rm km\,s^{-1}}\right)^{-1}
  \left(\frac{R_{\rm ej}}{6\times10^{13}\rm cm}\right)^{-2} 
\end{equation}
(see Eq.~(\ref{dens_z})).
 
We study a parameter space for $n_{\rm ej}$ and $n_{\rm w}$  varying between $10^{-5}\le M_{\rm ej}/M_{\odot}\le 10^{-2}$ and $10^{-6}\le  \dot{M_{\rm w}}/M_{\odot}\,\rm wk^{-1} \le 10^{-3}$. We consider $V_{\rm ej}=1000$\,km\,s$^{-1}$ and $3000$\,km\,s$^{-1}$. To classify the shocks we compare the cooling time given by Eq.~(\ref{tcool}) assuming free-free cooling with the typical duration of this phenomena $\sim 2$ weeks. The results are shown in Figure\,\ref{fig2}.

In the case of a slow ejecta with $V_{\rm ej} = 1000$\,km\,s$^{-1}$ the forward shock is always radiative, thus the nature of the shock system is given by the reverse shock, which is radiative for $M_{\rm ej} > 10^{-5.6}$\,M$_{\odot}$ (pink region on the left plot). The case of interest in this work, an adiabatic reverse shock with a radiative forward shock, corresponds to the smaller green region. In the case of a faster ejecta, $V_{\rm ej} = 3000$\,km\,s$^{-1}$, we obtain a similar plot as in the case of YSO jets in Fig.~\ref{fig1}. The area of the right plot in Figure\,\ref{fig2} is divided in: both adiabatic in the lower left (yellow), both radiative in the upper right (pink), an adiabatic forward shock and a radiative reverse shock in the top left (gray) and the opposite situation in the bottom right (green). In this case the condition for a reverse shock to be adiabatic with an efficiently cooling forward shock is the dominant in the given parameter space.   

The very high densities in novae make the shocks very prone to efficiently cool radiatively. However we can conclude that under the model adopted, the possibility of having adiabatic shocks in these systems is plausible, especially for $V_{\rm ej} > 1000\,$km\,s$^{-1}$. 

%%%%%%%%%%%%%%%%%%%%%%%%%%%%%%%%%%%%%%%%%%%%%%%%%%%%%%%%%%%%%%%%%%%%%%%%%
\section{Instabilities at the contact discontinuity}\label{instabilities}

The density of the plasma  downstream of the radiative forward (bow) shock is 
\begin{equation}\label{n_rad}
\frac{n_{\rm a}^{\prime}}{n_{\rm a}} = 233
\left(\frac{v_{\rm fs}}{100\,\rm km\,s^{-1}}\right)^2
\left(\frac{T}{10^4\rm K}\right)^{-1}
\end{equation}
\cite[e.g.][]{Blondin_90} when the plasma is cooled down to a temperature $T$, making the density contrast at the contact discontinuity $4 n_{\rm j}/n_{\rm a}^{\prime} \ll 1$. As a consequence, the contact discontinuity is unstable to dynamical and thermal instabilities. 
A dense layer of density $n_{\rm a}^{\prime}$ located at distance $l_{\rm th}$ downstream of the bow shock  fragments into several clumps \cite[e.g.][]{Calderon_20}. However, we note that the component of the magnetic field in the ambient medium parallel to the bow shock front ($B_{\rm a,\perp}$) limits the compression factor to a maximum value 
\begin{equation}\label{B_mc}
\frac{n_{\rm max}^{\prime}}{n_{\rm a}} \sim 78
\left(\frac{n_{\rm a}}{10^4\,\rm cm^{-3}}\right)^{\frac{1}{2}}
\left(\frac{v_{\rm fs}}{100\,\rm km\,s^{-1}}\right)
\left(\frac{B_{\rm a,\perp}}{0.1\rm mG}\right)^{-1}
\end{equation}
\citep{Blondin_90}. 

This indicates that significant enhancement in the plasma density downstream of the reverse shock is feasible if  instabilities grow fast enough to fragment the dense shell and form such clumps. 

\subsection{Rayleigh-Taylor instabilities}

The Rayleigh-Taylor (RT) instability can grow in the contact discontinuity due to the velocity shear and the force exerted by the downstream material of the reverse shock on the forward shock. 

If the forward shock is radiative, the formation of a shell much denser than $n_{\rm j}$ and $n_{\rm a}$ at the contact discontinuity makes the working surface unstable even in the case of light jets. Following the analysis in \cite{Blondin_90}, the acceleration of the dense shell with a width $W_{\rm sh}$ can be written as 
$a\sim  n_{\rm j} v_{\rm rs}^2 /n^{\prime}_{\rm a}W_{\rm sh}$.

The growth time of RT instabilities is $t_{\rm RT}\sim 1/\sqrt{a k}$, where $k$
is the wavenumber. By considering the  characteristic dynamical timescale $t_{\rm dyn} = 4 R_{\rm j}/v_{\rm j}$ we obtain
\begin{equation}\label{tau_RT}
\frac{t_{\rm RT}}{t_{\rm dyn}}\sim 
\left(\frac{\sqrt{\chi}}{4\sqrt{\chi}+1}\right)
\left(\frac{\lambda}{R_{\rm j}}\right)^\frac{1}{2}
\left(\frac{W_{\rm sh}}{R_{\rm j}}\right)^\frac{1}{2}
\left(\frac{T}{10^4\rm K}\right)^{-\frac{1}{2}}
\left(\frac{v_{\rm j}}{1000\,\rm km\,s^{-1}}\right).
\end{equation}
The condition $t_{\rm RT}/t_{\rm dyn}<1$ leads to 
\begin{equation}\label{tau_RT1}
\left(\frac{v_{\rm j}}{1000\,\rm km\,s^{-1}}\right) >
\left(\frac{4\sqrt{\chi}+1}{\sqrt{\chi}}\right)
\left(\frac{W_{\rm sh}}{R_{\rm j}/3}\right)^{-1}
\left(\frac{T}{10^4\rm K}\right)^{\frac{1}{2}}
\end{equation}
where we have assumed $\lambda=2 \pi/k \sim W_{\rm sh}$. 

\section{Numerical study}
\label{numerical}

We performed 2D hydrodynamic simulations with the PLUTO code \citep{2007ApJS..170..228M} to illustrate the physical processes mentioned in the previous sections. We are interested in simulating the physics in the interaction of the incoming material, shocked incoming material, contact discontinuity, target shocked material, and target material. Therefore we study the fluid collisions in a 2D rectangular box of size $L_{\rm x}\times L_{\rm y}$, with $L_{\rm x} = 4$\,$R_{\rm j}$ and $L_{\rm y} = 8$\,$R_{\rm j}$, in a uniform Cartesian grid of resolution 1024$\times$2048. Boundary conditions are periodic in the $x-$direction (horizontal) and outflow in the $y-$direction (vertical). The fluids are assumed to follow an ideal equation of state with an adiabatic index $\gamma = 5/3$.
 
In the presence of cooling we use the tabulated cooling function that includes metal-line cooling, see Section\,\ref{jetshocks}, from \citet[][]{2009A&A...508..751S}. Calculations are performed using a HLLC solver with parabolic reconstruction. The time integration is performed using a Runge-Kutta 2 algorithm, controlled by a Courant-Friedrichs-Lewy number of 0.4. 

We simulate the case of a protostellar jet ($R_{\rm j} = 10^{16}$~cm) impinging upon an ambient molecular cloud. The results can be extrapolated to the case of novae. Initially we have a fluid of density $n_{\rm j}= 5$~cm$^{-3}$ and velocity $v_{\rm j} = 300$\,km\,s$^{-1}$ $\hat{\textbf{\j}}$ at $T_{\rm j} = 1000$\,K for $y \leq 4\,R_{\rm j}$, colliding with a material at rest of density $n_{\rm a} = 500$\,cm$^{-3}$ and temperature at $T_{\rm a} = 100$\,K for $y > 4\,R_{\rm j}$, with $n_{\rm a}/n_{\rm j} = 100$. According to our analytical estimates in Sec.~\ref{jetshocks} this configuration should develop a working surface composed of an adiabatic reverse shock and a radiative forward shock. The white dot in Figure\,\ref{fig1} shows the position of these parameters in the shock diagnostic map.  
  
\subsection{Results}\label{results}

\begin{figure}
\centering
\includegraphics[width=0.49\textwidth,trim=0cm 0cm 0cm 0cm, clip=true,angle=0]{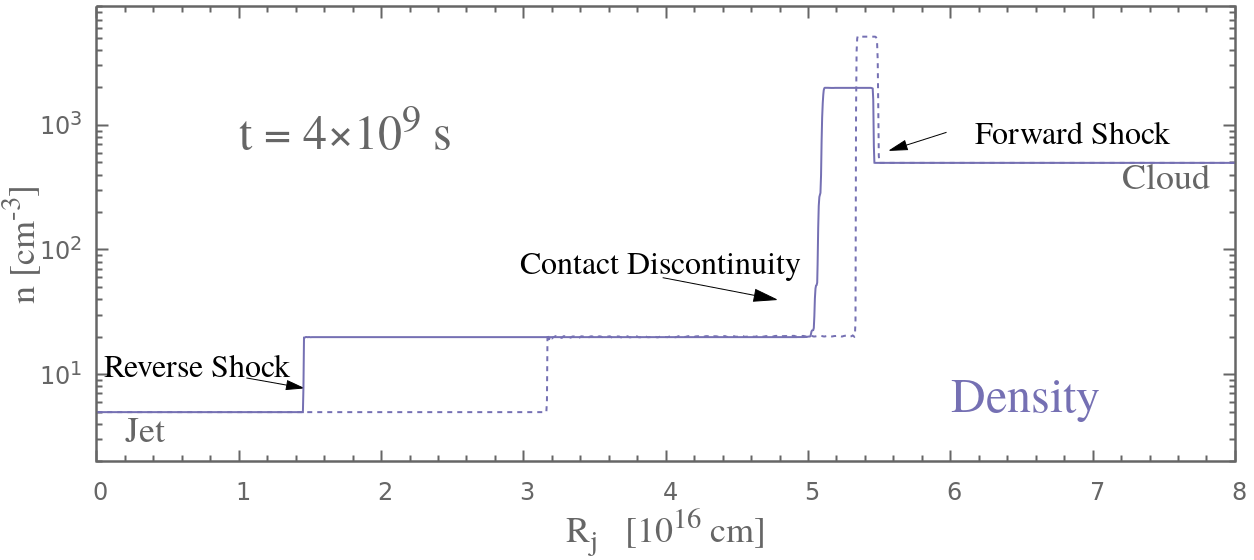}\\
\includegraphics[width=0.49\textwidth,trim=0cm 0cm 0cm 0cm, clip=true,angle=0]{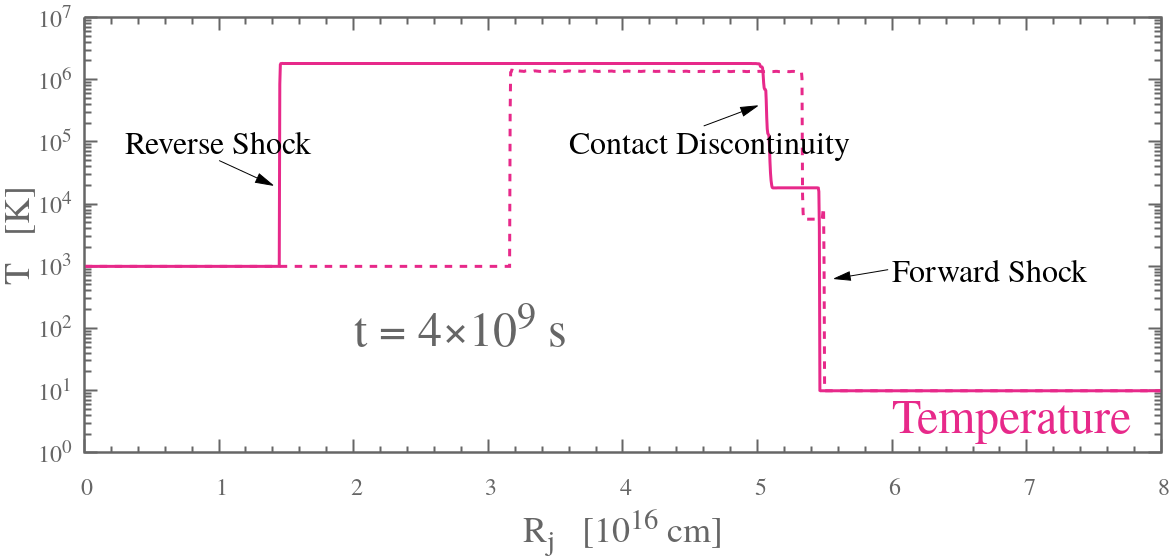}
\caption{Density (top) and temperature (bottom) profile along the vertical $y-$direction averaged in the horizontal $x-$direction for $t = 4\times10^{9}$\,s. Dashed and solid lines correspond to the cases with and  without cooling, respectively.}
\label{S1}
\end{figure}

Figure\,\ref{S1}  shows the density (top) and temperature (bottom) profile of the system along the $y-$direction, averaged in $x$, for a time $t = 0.04$\,$t_{\rm phy}$, where $t_{\rm phy} \equiv R_{\rm j}/10^{5}$\,cm\,s$^{-1} = 10^{11}$~s { is the physical time of the simulation}. Both shocks are strong, producing a density jump of the order of 4. The jet and ambient shocked gases reach temperatures of $1.8\times10^{6}$ and $1.8\times10^{5}$\,K, respectively. Using the postshock temperature, we measure the shock velocities in the simulations ($V_{\rm s} = 100 \sqrt{T_{\rm s} / 1.38  \times 10^5}$\,km\,s$^{-1}$), while calculating the displacement of the shock fronts as time evolves. The shock front cells are established with good accuracy by searching for the position of a strong gradient in the temperature profile. The velocity of the contact discontinuity is the fluid velocity of the shocked material, directly extracted from the simulation by measuring the velocities of the shocked jet and the ambient. The measured velocities of the shocks and the contact discontinuity are in very good agreement with the values predicted by the theory. Using the equations in Sec.~\ref{jetshocks} we find that $v_{\rm rs} = -362.72$~km~s$^{-1}$ and $v_{\rm bs} = 36.27$\,km\,s$^{-1}$. The contact discontinuity is expected to move with velocity  $v_{\rm cd} \approx 3 v_{\rm bs}/4 =27.27$\,km\,s$^{-1}$. 

In the presence of cooling the system acts as predicted theoretically: the shocked jet material experiences negligible cooling, while the shocked ambient material cools down. From the density (upper panel) and temperature (lower panel) plots  in  Figure\,\ref{S1} (dashed lines) we see the typical profile of a radiative forward shock. The material behind the forward shock cools, i.e. the temperature drops, and compresses reaching values of $n_{\rm a}^{\prime}\gg 4n_{\rm a}$. Even when the radiative losses do not affect the reverse shock dynamics, it can be noticed a small drop on the temperature { with $T_{\rm s, cool}/T_{\rm s} \sim 0.8$. The cooling of the forward shock changes the contact discontinuity velocity (see the displacement of the contact discontinuity in Figure\,\ref{S1} with respect to the case without cooling). This also slightly slows the reverse shock down, changing the shock Mach number $\mathcal{M}$ and hence the jump conditions across the shock \citep[see,  e.g.,][]{Antoine-21}. The reverse shock velocity measured in the simulation with cooling is $v_{\rm rs} \sim -324$\,km\,s$^{-1}$, giving $\mathcal{M}_{\rm cool} \sim 87.5$; in the case without cooling  $\mathcal{M} \sim 98$. The temperature jump condition can be written as $T_{\rm s}/T \sim 5/16\,\mathcal{M}^2$, for a strong shock with $\gamma = 5/3$. Therefore $T_{\rm s, cool}/T_{\rm s} \approx \mathcal{M}_{\rm cool}^2/\mathcal{M}^2 \sim 0.8$. A difference in the reverse postshock density is not observed and also not expected because for strong shocks the density jump condition is practically independent of the shock Mach number.}

\begin{figure}
\centering
\includegraphics[scale=0.4,trim= 15cm 0cm 12cm 0cm, clip=true,angle=270]{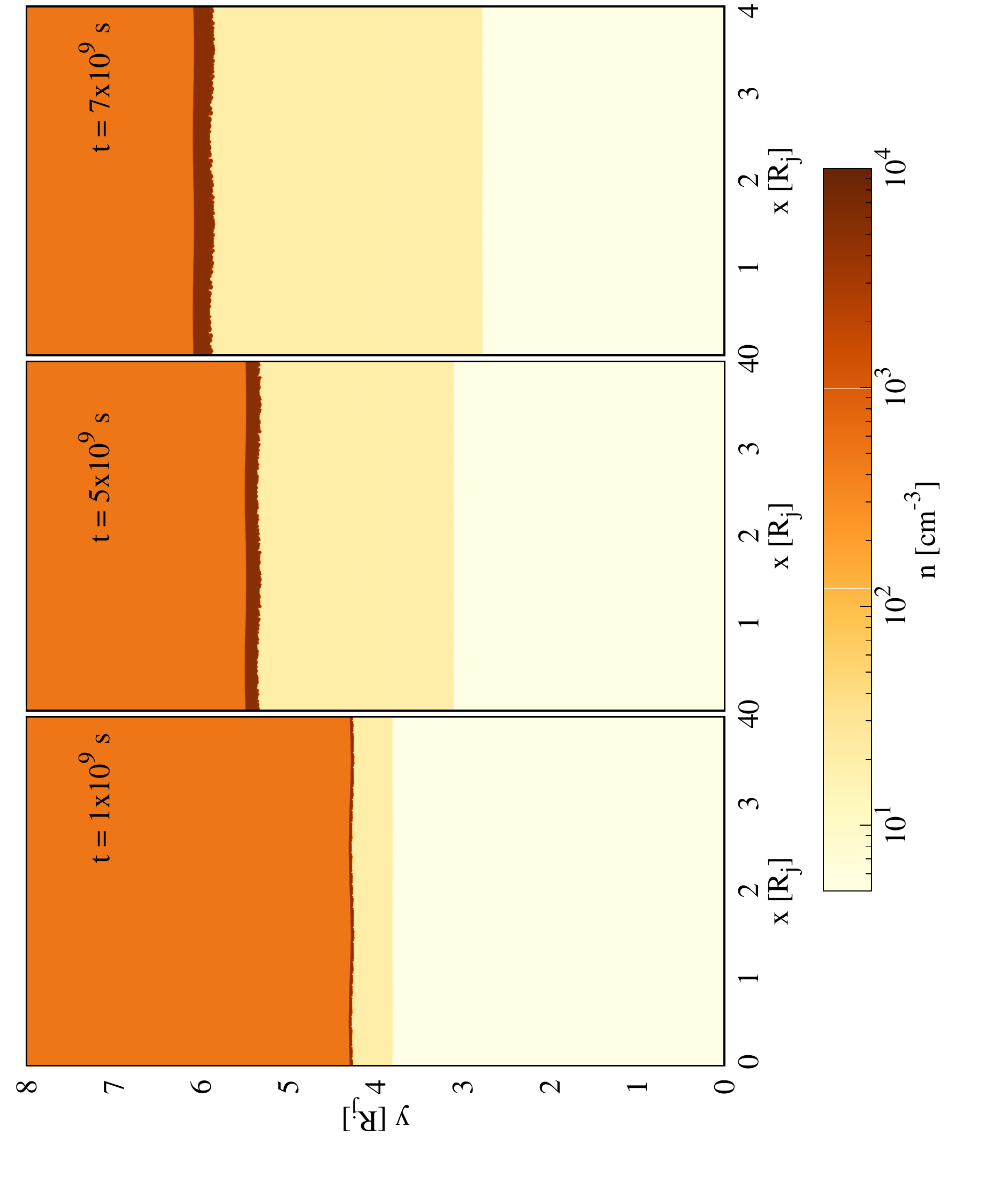}
\caption{Density maps for different evolving times. The interface is slightly perturbed (sinusoidal interface).}
\label{S3}
\end{figure}

The fluid profile propagates without disturbances in this simple configuration, without mixing of the jet and cloud materials. However, in a real situation the system will suffer perturbations produced, for example, by local inhomogeneities. In order to study the instabilities that might arise in the evolution of the system (see previous section), we consider a sinusoidal interface $y = y_{0} + 0.01 \sin(\frac{2\pi}{L}y)$, with $y_{0} = 4$ and $L = 2$ between the jet and cloud material, which mimics a perturbation\footnote{In a real situation perturbations are the norm arising for example from density inhomogeneities, velocity gradients, and irregular geometry among other effects.}. 
Figure\,\ref{S3} shows a sequence of density maps as times evolves. Instabilities develop in the contact discontinuity as predicted by the theory, which induce mixing and turbulence. The instability also heats the shocked ambient locally. The density structures in the unstable layer resemble the typical finger structures developed when the RT instability is operating.  

\begin{figure}
\centering
\includegraphics[width=0.14\textwidth,trim=5.5cm 0.5cm 5.5cm 0cm, clip=true,angle=270]{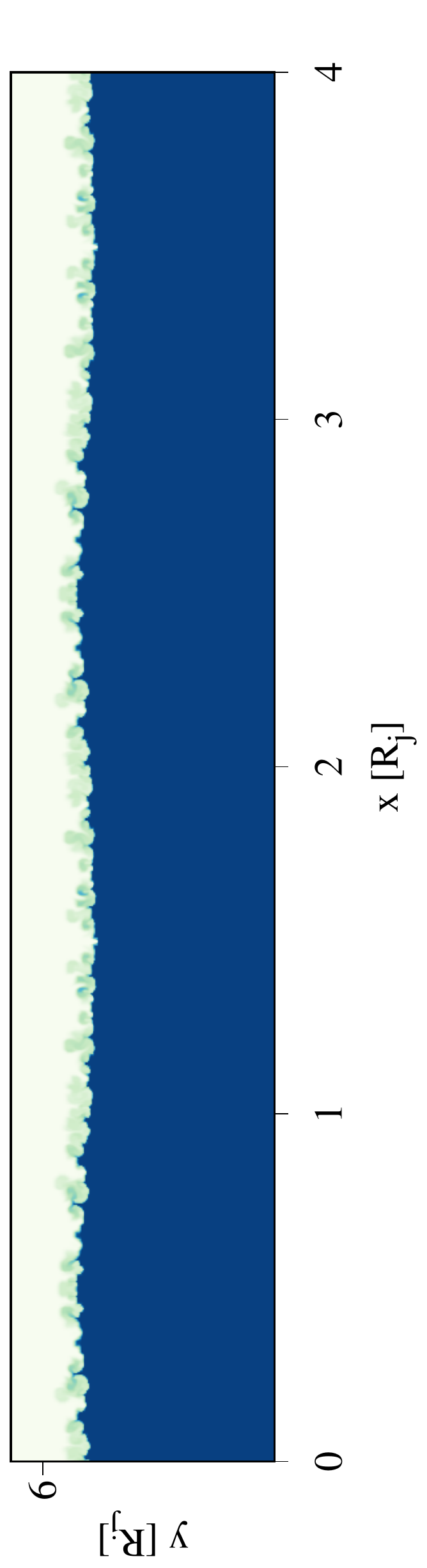}
%trim= arriba  izquierda abajo derecha
\caption{Tracer map at $t = 7\times 10^{9}$\,s, zoomed at the forward shock region. The tracer indicates the advection of the  jet and ambient materials as the system evolves. Initially the value 1 is designated to the jet material (blue) and 0 to the ambient one (white).}
\label{S5}
\end{figure}

In order to analyse the mixing of the two materials. we include a tracer field in the simulation that is advected with the fluid. Initially, the jet and ambient material have a tracer values of 1 and 0, respectively. Figure\,\ref{S5} shows the tracer map at $t = 7\times10^{9}$\,s, zoomed on the forward shock region. The denser shocked ambient material penetrates the shocked jet material, and it is clear that mixing occurs as expected.  

\begin{figure}
\centering
\includegraphics[width=0.4\textwidth,trim=0cm 1.5cm 0.5cm 1.5cm, clip=true,angle=270]{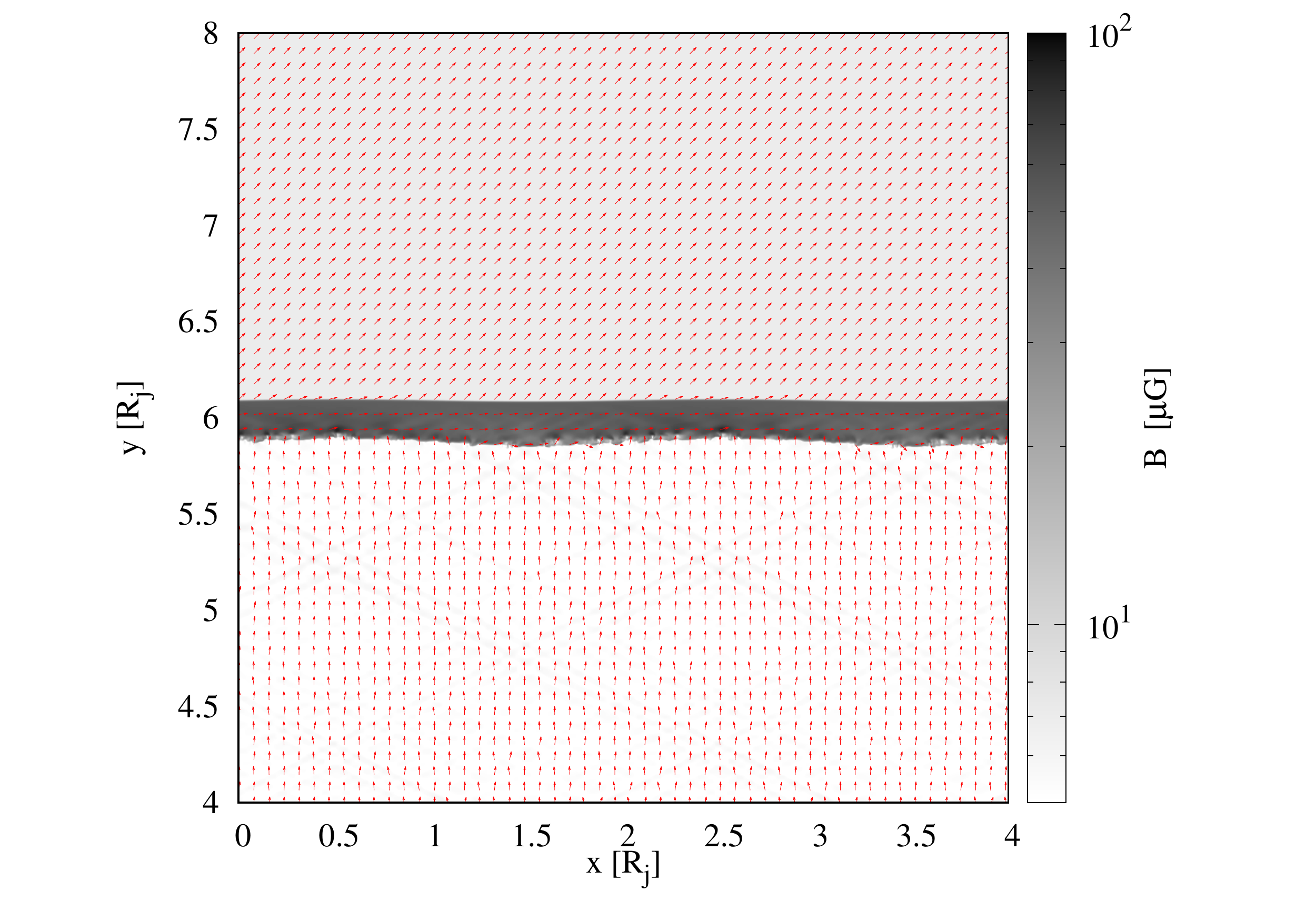}
%trim= arriba  izquierda abajo derecha
\caption{Magnetic field intensity map at $t = 7\times10^{9}$\,s. The red arrows show the magnetic field direction in the given grid point.}
\label{M1}
\end{figure}

\subsubsection{Magnetic field}

In order to illustrate the effects of a magnetic field we include a field $B_{\rm j} = 5$\,$\mu$G\,$\hat{\jmath}$ and $B_{\rm a} = 5\,\mu {\rm G}\,\hat{\imath} + 5\,\mu {\rm G}\,\hat{\jmath}$ in the MHD simulation setup. This gives a thermal to magnetic pressure ratio $\beta$ of 0.7 for the jet, and 3.5 for the ambient medium. We also consider the case in which the magnetic field has no components perpendicular to $V_{\rm sh}$ (no $\hat{\imath}$-component). The ambient magnetic field component parallel to the shock front  is compressed by the forward shock. Without cooling, the magnetic field compression is $\sim$4, the same factor as the density. In the presence of cooling, further compression is expected and the field is amplified by a factor of $\sim$8. The magnetic field map at $t = 7\times10^{9}$\,s is shown in Figure\,\ref{M1}. The red arrows indicate the magnetic field direction in each cell plotted over the magnetic field intensity.

\begin{figure}
\centering
\includegraphics[scale=0.4,trim=0cm 0cm 0cm 0cm, clip=true,angle=270]{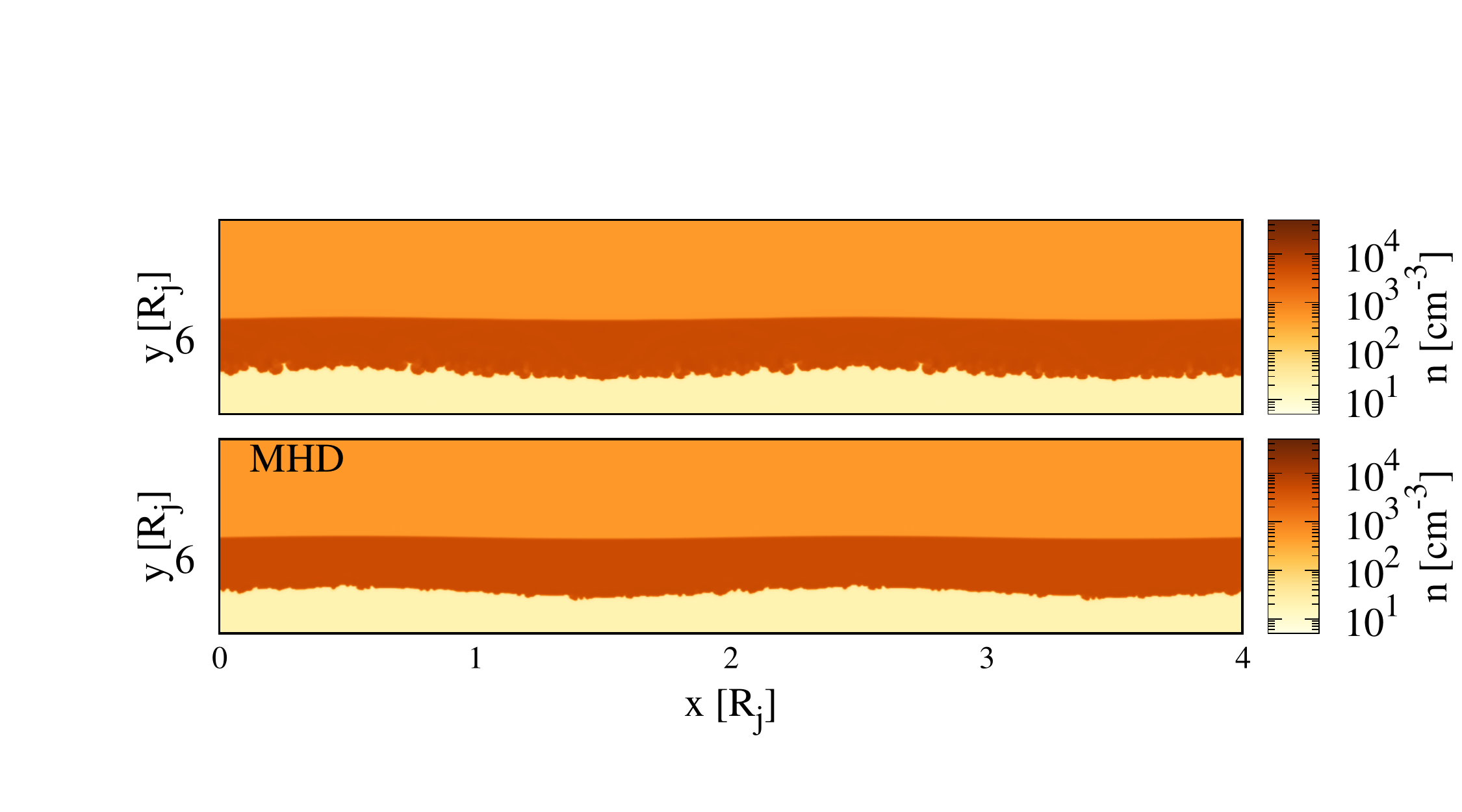}
%trim= arriba  izquierda abajo derecha
\caption{Density maps, zoomed at the forward shock region, at $t = 7\times10^{9}$\,s for the HD (top) and MHD (bottom) regime.}
\label{M2}
\end{figure}

The presence of a magnetic field inhibits or hinders the development of the instabilities discussed in Sect.\,\ref{instabilities}. The effect of the magnetic field in the unstable layer can be seen in Figure\,\ref{M2}, where we show the density map zoomed on the forward shock region at $t = 7\times10^{9}$\,s, for hydrodynamic (top) and MHD (bottom) scenarios. Although the presence of the magnetic field reduces the development of instabilities and  material mixing in the contact discontinuity, this effect is highly dependent on the magnetic field orientation. In the other case considered (only parallel $B$) the density structures developed by the instabilities exhibit a similar shape and level of material mixing as the non-magnetic case. In an actual astrophysical scenario, establishing the direction of the magnetic field is highly difficult and a number of assumptions are required.  

\subsection{Power Spectrum}

\begin{figure}
\centering
\includegraphics[scale=0.38,trim=0cm 2.8cm 0.cm 4cm, clip=true,angle=270]{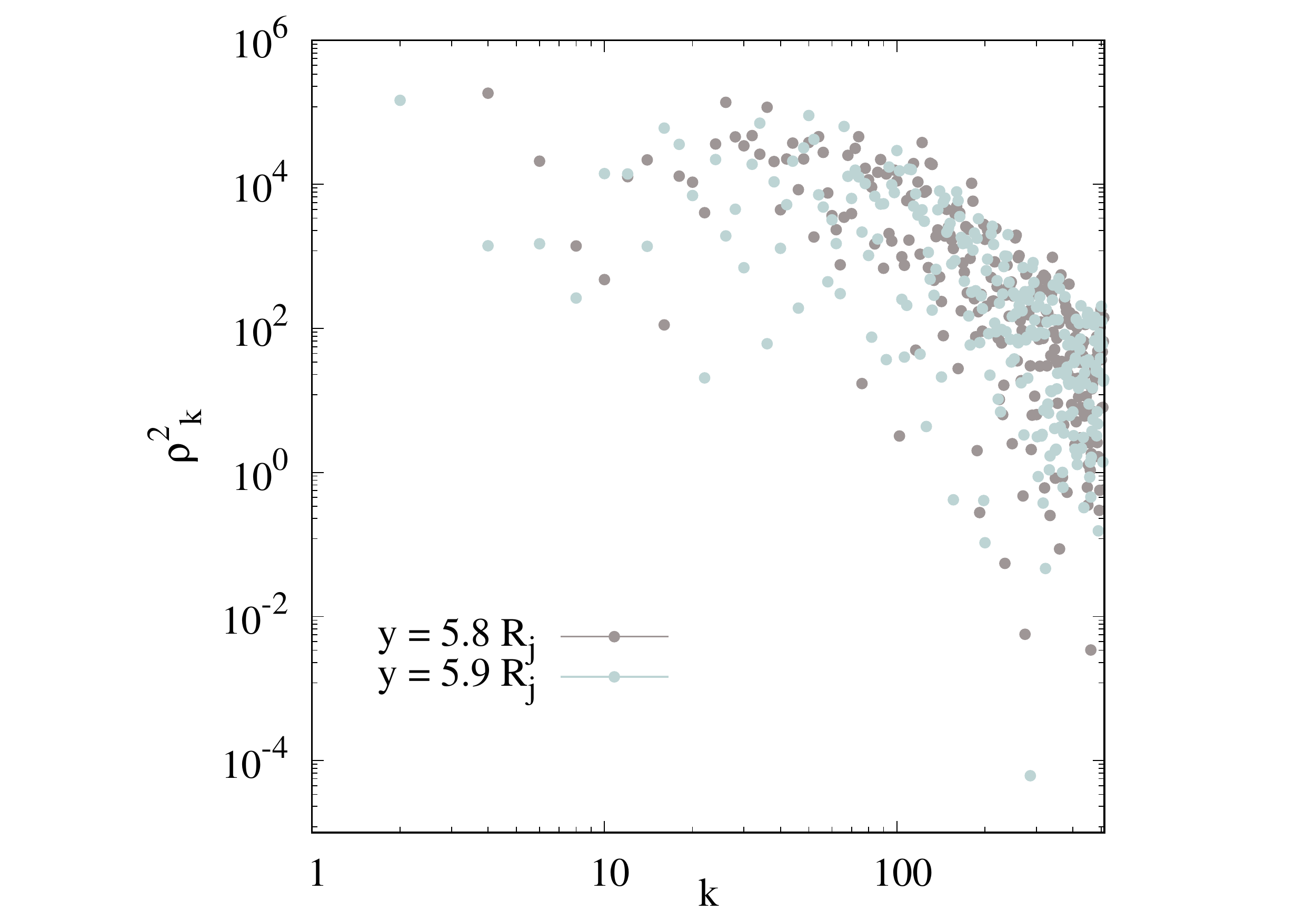}\\
\includegraphics[scale=0.38,trim=0cm 2.8cm 0.cm 4cm, clip=true,angle=270]{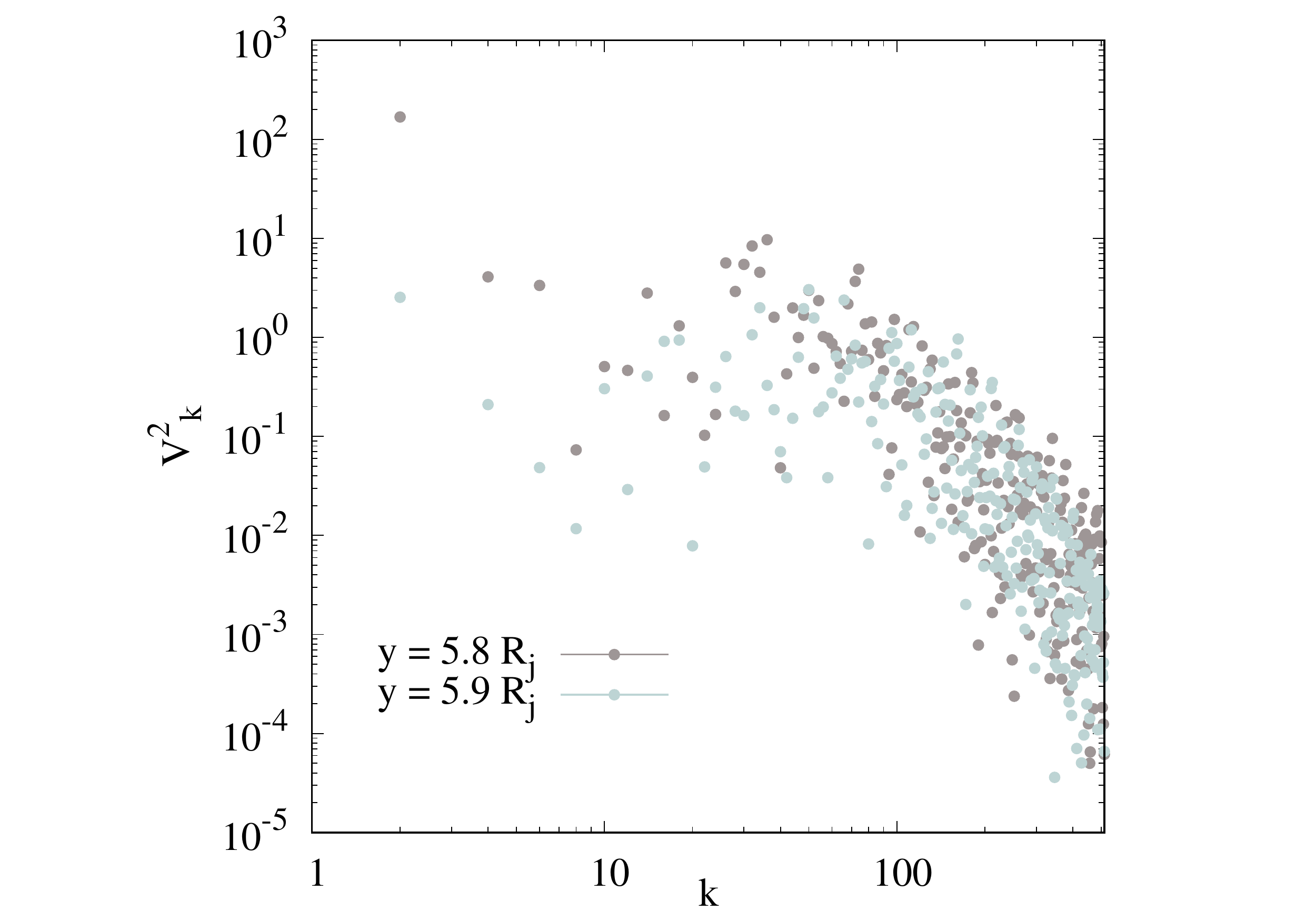}
%trim= arriba  izquierda abajo derecha
\caption{Density (top) and velocity (bottom) power spectrum at $t = 7\times 10^{9}$\,s for two different heights.}
\label{PS}
\end{figure}

We compute the power spectrum in the $x$ direction for fixed height $y$, at $t =7\times10^{9}$\,s. Figure~\ref{PS} shows power spectra for the density (top) and the velocity (bottom) for two values of $y$. These heights correspond to different regions where density perturbations appear due to instabilities (see upper plot of Fig.~\ref{M2}). The mixing coefficient  { is defined as} $\mathbf{D_{\rm mix} =  \sqrt{k_{\rm max} V^{2}_{k_{\rm max}}}k^{-1}_{\rm max}}$, where $k_{\rm max}$ is the scale of the fastest growth rate of the unstable mode in the velocity. From Fig.\,\ref{M2} we  estimate $k_{\rm max}\sim 30$ and $V^{2}_{k_{\rm max}}\sim 8.5$. { In physical units this yields  $D_{\rm mix} \sim 2.4 \times 10^{21}$\,cm$^2$\,s$^{-1}$}. We can estimate the mixing time for a given size $L_{0}$ by computing

\begin{equation}\label{t_mix}
t_{\rm mix} = \frac{L_0^2}{D_{\rm mix}} \sim 4.2\times 10^{10}
\left(\frac{L_0}{R_{\rm j}}\right)^2
\left(\frac{D_{\rm mix}}{2.4\times10^{21}\, \rm cm^2\,s^{-1}}\right)^{-1}\,{\rm s}.
\end{equation}

In the next section we discuss the relevance of efficient mixing for enhancement of the gamma-ray emission by the interaction of relativistic particles with matter fields.

\section{Gamma-ray emission}\label{discussion}

A system of two shocks: one radiative and the other adiabatic can be present in YSO jets and novae outflows, as it was demonstrated in Section~\ref{jetshocks}. The fast reverse shock is more efficient for accelerating particles, as the acceleration time for electrons and protons with energy $E_e$ and $E_p$, respectively, is
\begin{equation}\label{t_acc}
\frac{t_{\rm acc}}{\rm s} \sim 2.4\times10^7 
\left(\frac{E_{e,p}}{\rm TeV}\right)
\left(\frac{B}{\rm mG}\right)^{-1}
\left(\frac{v_{\rm sh}}{1000\, \rm km\,s^{-1}}\right)^{-2}
\end{equation}
in the Bohm diffusion regime. 

Furthermore, the adiabatic reverse shock has a  luminosity
\begin{equation}\label{L_sh}
\frac{L_{\rm sh}}{\rm erg\,s^{-1}} \sim 3\times10^{35} 
\left(\frac{R_{\rm j}}{10^{16}\rm cm}\right)^2
\left(\frac{n}{10^3\rm cm^{-3}}\right)
\left(\frac{v_{\rm sh}}{1000\, \rm km\,s^{-1}}\right)^{3}
\end{equation}
which is higher than the forward radiative shock and can power the shock-accelerated population of particles. Also, the lower densities in the jet in the case of YSOs could avoid injection problems produced by ionization and collision losses \citep[e.g.,][]{1996A&A...309.1002O}. 
 
Accelerated electrons and protons are injected in the shock downstream region with a luminosity $L_{e,p} = f L_{\rm sh}$, where $f\sim 0.1$ and 0.005 for the case of adiabatic and radiative shocks, respectively \citep{Caprioli_14a}. These particles must  reach the contact discontinuity in order to interact with the cooled compressed layer. In the transport of the particles various physical ingredients play a role. We consider for simplicity only the spatial diffusion (energy dependent) and the advection by the large scale gas velocity (energy independent). We only consider Bohm diffusion in the shock acceleration process, however far from the shock we expect a faster diffusion regime. Assuming a diffusion coefficient of  $D = 10^{25} \left(E_{e,p} / 10\,{\rm GeV} \right)^{0.5}$\,cm$^{2}$\,s$^{-1}$ (slower than the typical value in the ISM due to the instabilities) we obtain 
\begin{equation}\label{t_diff}
\frac{t_{\rm diff}}{\rm s} \sim 10^{7}
\left(\frac{R_{\rm j}}{10^{16}\, \rm cm}\right)^2
%\left(\frac{D_{10}}{10^{25}\, \rm cm^{2}\,s^{-1}}\right)^{-1}
\left(\frac{E_{e,p}}{10\,\rm GeV}\right)^{-\frac{1}{2}}.
\end{equation}
Advection downstream of the adiabatic shock over a distance $R_{\rm j}$ can be written as
\begin{equation}\label{t_adv}
\frac{t_{\rm adv}}{\rm s} \sim 4\times10^{8}
\left(\frac{R_{\rm j}}{10^{16}\, \rm cm}\right)
\left(\frac{v_{\rm sh}}{1000\, \rm km\,s^{-1}}\right)^{-1}.
\end{equation}
We define the residence time of particles in the downstream region as 
$T = \min\{{t_{\rm adv},t_{\rm diff}}\}$. 

The instabilities produced in the cooled layer/non-cooled shocked material (see e.g., Fig.\,\ref{M2}) act to facilitate the interaction of particles with the denser, cool material. In addition, the mixing facilitates the transport and the collision of particles with denser material radiating more non-thermal emission via pp collisions (for hadrons) and relativistic Bremsstrahlung (for leptons). The timescale is very similar in both cooling processes. The simplest form of the relativistic Bremsstrahlung and pp cooling time reads

\begin{equation}\label{t_Brem}
\frac{t_{\rm Br,pp}}{\rm s} \sim 2\times10^{11}
%\left(\frac{E_e}{\rm TeV}\right)^{-1}
\left(\frac{n}{10^4\,\rm cm^{-3}}\right)^{-1}.
\end{equation}
Note that $t_{\rm Br,pp}\propto n^{-1}$ and therefore cooling through pp inelastic collisions and relativistic Bremsstrahlung becomes very efficient in the cooling layer where the density is significantly larger than $4n_{\rm j}.$

The synchrotron cooling time of electrons in a magnetic field $B$ is
\begin{equation}\label{t_s}
\frac{t_{\rm syn}}{\rm s} \sim 4\times 10^8
\left(\frac{E_e}{\rm TeV}\right)^{-1}
\left(\frac{B}{\rm mG}\right)^{-2}.
\end{equation}
Inverse Compton (IC) scattering of IR/optical photons from the central object  with luminosity $L_{\star}$ and  energy density $U_{\rm ph}=L_{\star}/(4\pi z^2 c)$ has a characteristic timescale
\begin{equation}
    {t_{\rm IC} \over {\rm s}} %= \frac{
    \simeq 
    1.6\times 10^{10}
    \left(\frac{E_e}{\rm TeV}\right)^{-1}
    \left(\frac{L_{\star}}{10^4 L_{\odot}}\right)^{-1}
\left(\frac{z}{10^{17}\rm cm}\right)^{2},
\end{equation}
where $z$ is the distance from the photon source.

\subsection{Young stellar objects}

Gamma-ray emission from YSO jets has been modelled in some recent papers \cite[e.g.][]{Araudo_21}, but its detection has not been claimed to date. We show here that efficient mixing by RT instabilities in the contact discontinuity can significantly enhance the predicted gamma-ray emission levels, making them detectable in the GeV domain.

\begin{figure}
\centering
\includegraphics[scale=0.6,trim=0cm 0cm 0.cm 0cm, clip=true,angle=0]{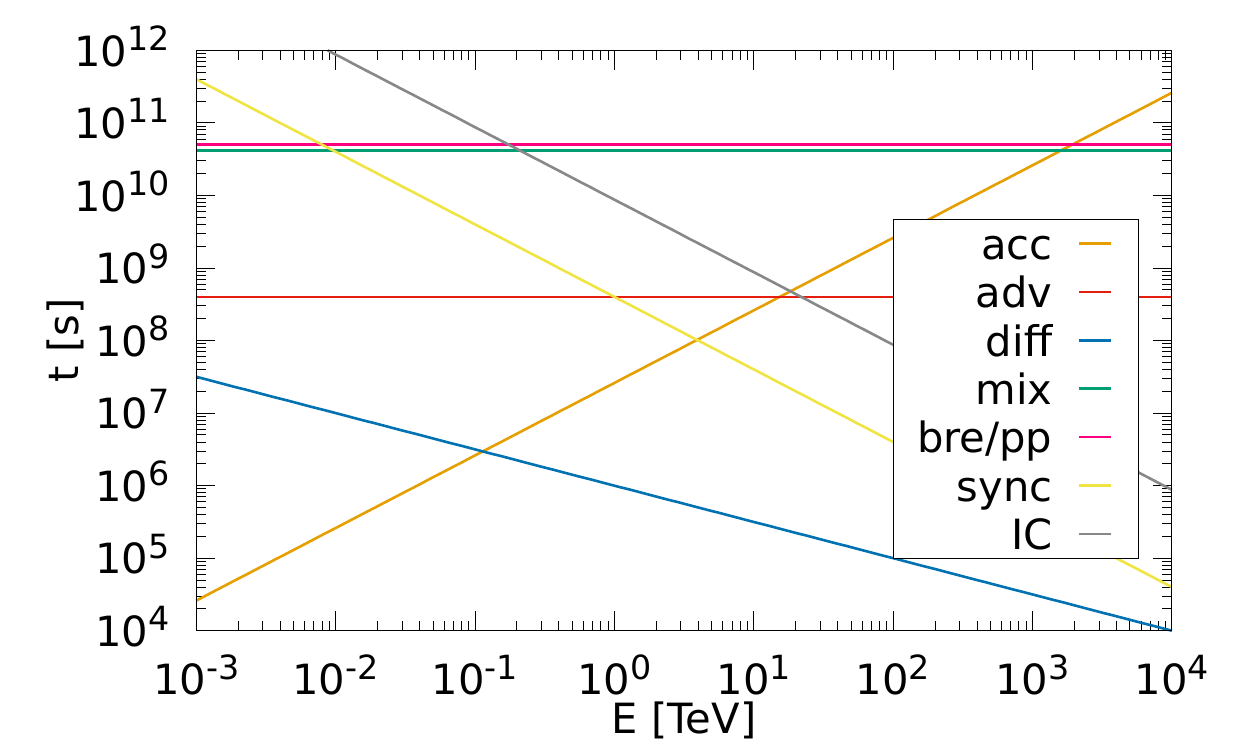}
%trim= arriba  izquierda abajo derecha
\caption{Time scales as a function of energy for the interaction of high-energy particles for a YSO.}
\label{timescales}
\end{figure}

By considering $B = 1$~mG, $V_{\rm sh} \sim 1000\,$km\,s$^{-1}$ and a typical length scale $R_{\rm j} \sim 10^{16}$\,cm we plot in Figure\,\ref{timescales} the timescales of the above mentioned processes. We also plot $t_{\rm mix}$ for comparison only, given that this value was not computed at the steady state. We can see that the acceleration is very efficient, together with the transport of particles by diffusion, giving a maximum energy of both electrons and protons of about $100$~GeV. This transport ensures that the accelerated particles in the reverse shock reach the denser regions where materials start to mix and nonthermal emission is enhanced. On the contrary, the transport by advection drag the particles away, downstream the shocked jet material which is subdominant in this case. 

Even though the real scenario might be much more complicated, these timescales indicate the dominating processes. We note however that if the magnetic field is locally amplified by non-resonant hybrid instabilities, the maximum energy of protons will be probably determined by the available amplification time \citep{Araudo_21}. A detailed model is beyond the scopes of the current study, but will be presented in a future work.

\subsection{Novae}

Gamma-ray emission has been detected from more than a dozen novae in the GeV-range with most of the sources being classical novae \cite[see][for a recent review]{2020arXiv201108751C}. For example, {\it Fermi} and H.E.S.S. have recently detected HE and VHE gamma-ray emission from the recurrent nova RS Ophiuchi (ATel\#14834 and \#14857, respectively); see further details in Sect.\,\ref{rsophi}. The gamma-ray emission spans several orders of magnitude among the detected sources \citep{2018A&A...609A.120F}. This difference might arise simply because the physical parameters in each source are slightly different. For example, we can see from the shock diagnostic maps in Figure\,\ref{fig2} that a change in a factor of 3 in velocity changes radically the possibilities of having adiabatic shocks in the system. Furthermore, changes in metallicity not considered here can modify the cooling function and might change the shock radiative efficiency. 

A correlation between the optical and the gamma ray emission has been claimed \citep[e.g.,][]{2017NatAs...1..697L,2020NatAs...4..776A}. This correlation appears to be strong in some systems but does not behave equally in all detected sources. In the systems where a correlation exists, it is highly probable that the emission is coming from the same spatial region, i.e. a radiative shock. However this is not in conflict with our claims. Firstly, not all the possible physical parameters in novae result in adiabatic shocks and some sources might have an adiabatic-radiative shock combination.  Secondly, even if the optical and gamma emission are coming from the same radiative shock, this does not confirm that the emitting nonthermal particles have been accelerated in that same shock. In fact, the emission might arise when a underlying population of relativistic particles, accelerated in the reverse shock for example, is enhanced by the strong radiative shock compression (responsible for the optical emission) and even suffer reacceleration \citep[e.g.,][]{1982ApJ...260..625B}. An interesting case that supports our findings is that of the nova V959 Mon \citep[e.g.,][]{2012CBET.3202....1F}. This source has been detected as a GeV gamma-ray transient by {\it Fermi} \citep{2014Sci...345..554A}. X-ray data analysis indicates that the reverse shock should be non-radiative \cite[see,][]{2021MNRAS.500.2798N}.

The hadronic scenario is the one of the most favourable for explaining the high-energy radiation \citep[e.g.,][]{2017NatAs...1..697L,2018A&A...612A..38M}. In this scenario, for protons to produce a gamma-ray of energy $E$ through proton-proton collisions, they need to have energies at least ten times $E$, which is not the case for electrons emitting through relativistic Bresmsstrahlung. This last case favours the scenario of particles being accelerated at a reverse shock and radiating elsewhere, given that the most energetic particles diffuse more efficiently. A detailed modeling is needed to quantify the viability of the adiabatic-radiative shock scenario and this will be addressed in future works. Below we analyse the case of RS Oph, recently detected at gamma-rays.

\subsubsection{The case of RS Oph}\label{rsophi}

RS Oph is a symbiotic recurrent nova system that undergoes thermonuclear outbursts  approximately every 20 years \citep[e.g,][]{2008ASPC..401...31A}. The last detected optical outburst was during August 2021 (vsnet-alert 26131\footnote{\url{http://ooruri.kusastro.kyoto-u.ac.jp/mailarchive/vsnet-alert/26131}}). The binary system, located at $d = 1.6\,$kpc \citep{1987rorn.conf..241B}, is composed by a white dwarf and a red giant (RG) companion \citep{1994AJ....108.2259D}. { Here we assume that in} this source the shocks are produced in the collision of a fast wind with the dense and slow wind of the RG star \citep{2007ApJ...665..654V,2011ApJ...740....5V}. { However, other models for RS Oph exist in the literature. In particular, given that the white dwarf in RS Oph is close to the Chandrasekhar limit some authors have modeled the system similarly to a supernova expanding in a wind medium \citep[e.g.,][]{2008A&A...484L...9W,2016MNRAS.457..822B}.}

The fast wind has an inferred velocity $V_{\rm w} > 6000$\,km\,s$^{-1}$ and a mass loss-rate  $\dot M_{\rm w} \sim 1.6\times10^{-5}\,{\rm M}_{\odot}\,{\rm yr}^{-1}$ \citep{2011ApJ...740....5V}. The RG wind has a velocity of $\sim$ 15\,km\,s$^{-1}$ and  $\dot M_{\rm RG} = 2\times10^{-7}\,{\rm M}_{\odot}\,{\rm yr}^{-1}$. We assume a time scale $t_{\rm ej} \sim 2$ weeks as  in Sect.\,\ref{nova}, and an orbital separation $a = 1.48\,$AU \citep[e.g.,][]{10.1093/mnras/stw001}. For estimating the wind density $n_{\rm RG}$ at a distance $a$ we use Eq.\,(\ref{w}) properly normalized; we calculate the fast wind density as $n_{\rm w} = {3\dot{M}_{\rm w}t_{\rm ej}}/\left({4\pi m_{\rm p}(V_{\rm w}t_{\rm ej})^3}\right)$ . Using the expressions from Sect.\,\ref{jetshocks} we obtain $v_{\rm bs} \sim 130\,$km\,s$^{-1}$, $v_{\rm rs} \sim 5900\,$km\,s$^{-1}$ for $\chi \sim 5\times 10^{-4}$. We estimate $t_{\rm cool}$ (see Eq.\,\ref{tcool}) for the reverse shock, propagating through the fast wind, and the forward shock that develops in the RG wind in this case. We conclude that the reverse shock is highly adiabatic whereas the forward shock is highly radiative during $t_{\rm ej}$.

\begin{figure}
\centering
\includegraphics[scale=0.6,trim=0cm 0cm 0.cm 0cm, clip=true,angle=0]{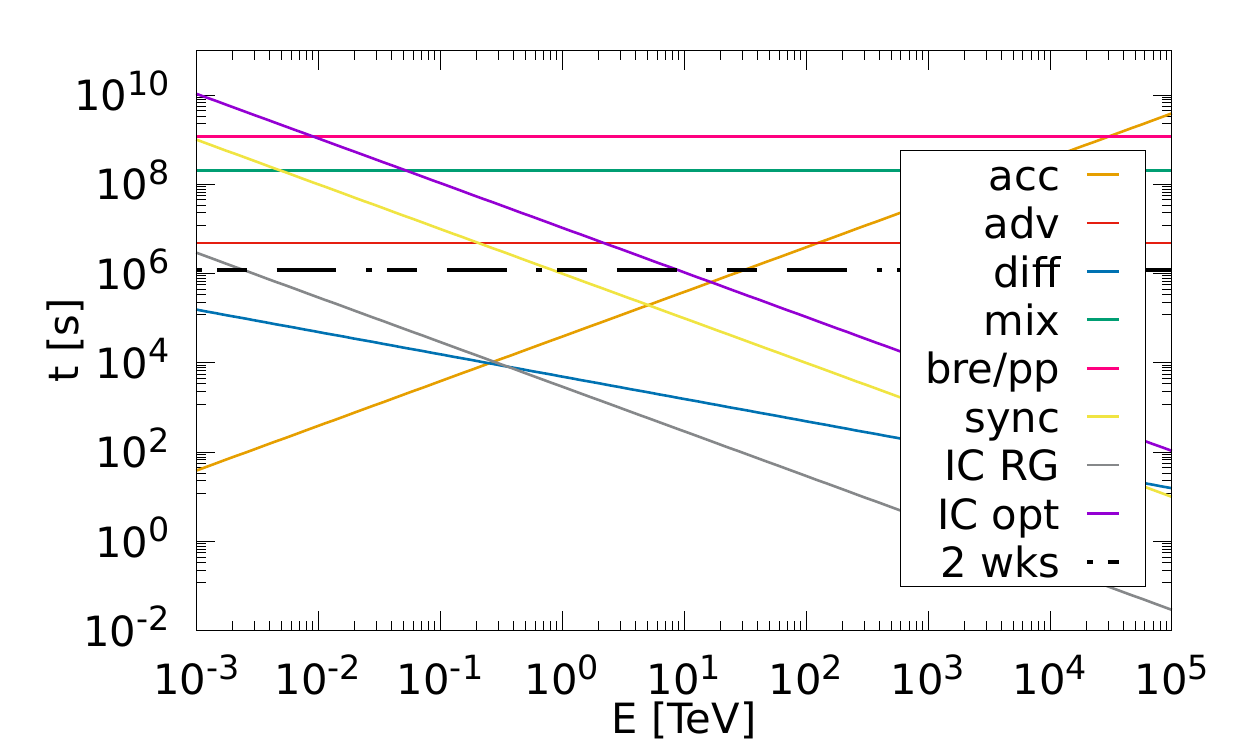}
%trim= arriba  izquierda abajo derecha
\caption{Idem as Fig.~\ref{timescales} but for the nova RS Oph.}
\label{timescales2}
\end{figure}

The magnetic field strength near the reverse shock, $B = 2\times 10^{-2}$\,G, is estimated assuming that the magnetic pressure is a fraction $\epsilon_{B} = 10^{-4}$ of the thermal pressure of the post-shock gas \citep[see][]{2015MNRAS.450.2739M}. The target radiation fields for IC in the vicinity of the reverse shock are the RG photon field $U_{\rm RG}$ at distance $\sim a$ and the optical $U_{\rm opt}$ emission from reprocessed X-rays \citep[][]{2014MNRAS.442..713M}. The companion star of the system is a M2III giant star \citep[][]{2018MNRAS.480.1363Z}, we estimate a luminosity $L_{\rm RG} \sim 2.5\times10^{36}\,$erg\,s$^{-1}$ (see the adopted stellar parameter is Table\,\ref{table:rs}). For estimating $L_{\rm opt} \sim 7\times10^{35}$\,erg\,s$^{-1}$ we assume that a small fraction, $10^{-2}$, of the shock power $L_{\rm sh}$ is radiated and reprocessed into optical emission.

In Figure\,\ref{timescales2} we show the relevant time scales involved in particle acceleration, diffusion and radiation loses in the reverse adiabatic shock. IC losses are very efficient giving electrons  maximum energies  $\sim$ 0.3\,TeV. In the case of protons, the losses for $pp$ do not affect the acceleration, which is limited only\footnote{We do not consider any effect from the neutrals that might affect particle acceleration \citep[see, e.g.,][]{2016MNRAS.457.1786M}.} by the time-scale of the event giving $E_{p,\rm max}\sim$ 30\, TeV. The diffusion of particles into the region of the contact discontinuity is fast, allowing the particles to further radiate there. The maximum energy of electrons and protons are compatible with high and very-high-energy gamma emission.

\begin{table}[t]
\caption{Model and inferred parameters for RS Oph.}
\begin{tabular}{ll}
\hline
Parameter & Value\\
 \hline
fast wind velocity $[V_{\rm w}]=\rm  km\,s^{-1}$ &  6000\\
RG wind velocity $[V_{\rm RGw}]=\rm  km\,s^{-1}$ &  15\\
fast wind mass-loss rate $[\dot{M}_{\rm w}] =\rm M_{\odot}\, wk^{-1}$ & $6\times10^{-7}$\\
RG wind mass-loss rate $[\dot{M}_{\rm RG}] =\rm M_{\odot}\, yr^{-1}$ & $2\times10^{-7}$\\
RG density $[n_{\rm RG}]=\rm  cm^{-3}$ &  $8.2\times10^8$\\
fast wind density $[n_{\rm w}]=\rm  cm^{-3}$ &  $4.4\times10^5$\\
RG temperature $[T_{\rm RG}] =\rm K$   &  3750 \\
RG radius $[R_{\rm RG}] =\rm cm$       &  $4.2\times10^{12}$\\ 
\hline
\label{table:rs}
\end{tabular}
\end{table}

RS Oph was detected for the first time at gamma rays on August 2021 by {\it Fermi} LAT operating from 20\,MeV to 300\,GeV; the satellite detected a transient gamma-ray source positionally consistent with the nova. For its part, H.E.S.S. which operates in the energy range 10\,GeV to 10\,TeV also detected a very-high-energy gamma-ray excess compatible with the direction of RS Oph. Following the detection, H.E.S.S. observations continued during the nova outburst. Here we show with a simple estimation that an adiabatic reverse shock is expected in RS Oph, and that it is capable of accelerating electrons and protons to high energies. The expected maximum energies are ~ 0.3\,TeV for electrons and ~ 30\,TeV for protons. In such dense environments we expect significant $pp$ emission up to 3 TeV. IC and Bremsstrahlung would be important at hundreds of GeV. Additional high-energy radiation, especially Bremsstrahlung and $pp$, is expected when particles diffuse to the contact discontinuity and compressed RG wind regions. The interaction of these high-energy particles can explain the gamma-ray emission detected. A detail model of the system will be possible when the observations are available.

\section{MHD scaling of YSO jets and novae to laboratory experiments}
\label{scaling}

In order to extend the scope of this work, we present preliminary estimates of the feasibility to scale the shocks from YSO jets and novae outflows to laboratory experiments. Laboratory experiments with dense, magnetised plasmas provide a novel approach to the study of astrophysical jets and outflows. The experiments are typically conducted on high-power laser and pulsed-power facilities \citep{Remington2006RevModPhys}, with each experimental approach allowing for different ranges of plasma parameters which can be chosen to match different regimes of interest physics-wise. Typically, the experiments are characterised by temperatures $\sim$100-1000s's eV, flow velocities $\sim$100-1000's km~s$^{-1}$, plasma volumes $\sim$mm-cm$^3$, timescales $\sim$1-100s ns, and electron densities $\gtrsim10^{18}$ cm$^{-3}$. The effect of magnetic fields can be controlled in the experiments depending of the way the plasma is driven in the experiment. In the case of pulsed-power generators the magnetic field is produced from the strong electrical currents that drive the plasma whereas, in the case of laser experiments, they can be generated from strong gradients of electron density and temperature in the plasma due to the Biermann battery effect, or added introduced externally using capacitor-coil targets \citep{Santos2018PhPl} or pulsed-power systems like MIFEDS \citep{Fiksel2015RScI}. 

MHD scaling arguments \citep[e.g.,][]{Ryutov1999ApJ, Ryutov2000ApJS, Falize2011ApJ, Cross2014ApJ} make it possible to study astrophysical processes through laboratory experiments, for instance the launching and propagation of YSO jets \citep{Lebedev2019RevModPhys} and accretion shocks \citep{VanBoxSomMNRAS2017}. Recently, the self-similar  dynamics of the collision between adiabatic and radiative supersonic flows has been studied by \cite{Antoine-21}, including an analysis of their scaling to laboratory experiments. Following the details in \citep{Ryutov1999ApJ}, the scaling is based on five characteristic physical parameters for the astrophysical and laboratory systems: length scale ($R$), density ($n$), pressure ($P$), velocity ($v$) and magnetic field ($B$). The ratio of length scales, density and pressure result in three arbitrary scaling factors $a$, $b$ and $c$ which are further combined to constrain the time scale ($t$) and magnetic field. 

The two systems will evolve identically if the initial conditions for the physical parameters are geometrically similar and two dimensionless parameters, the Euler number $Eu = v \sqrt{P/\rho}$ and thermal plasma beta $\beta = 8 \pi P / B^2$ are the same. The scaling will be valid if both systems have a fluid-like behaviour, i.e. a localization parameter $\delta\ll1$, and negligible dissipation processes, i.e. Reynolds ($Re$), Peclet ($Pe$) and magnetic Reynolds numbers ($Re_M$) $\gg1$. 

Table 2 summarizes the MHD scaling for YSO jets and novae to a possible laboratory experiment. We fix the input astrophysical parameters (e.g. based on Table\,\ref{Tab:0}) and propose sensible laboratory parameters that match the scaling, resulting in scaling factors $a=10^{17}$ and $b=c=10^{18}$. For both YSO jets and novae, the laboratory parameters result in lengths scales $R_{\rm lab}\sim1$ mm, densities $n_{\rm lab}\sim10^{19}$ cm$^{-3}$, pressures $P_{\rm lab}\sim10^{5}$ bar, velocities $v_{\rm lab}\sim100-700$ kms$^{-1}$, magnetic fields $B_{\rm lab}\sim1-10$ T, and time scales $t_{\rm lab}\sim1$ ns. The temperature $T_{\rm lab}\sim1$ keV was obtained under the assumption of astrophysical temperatures $T_{\rm astro}=50$ eV (5.8$\times10^5$ K) which is in line with those obtained in the simulations in Fig. 3. The dimensionless parameters in Table 2 fulfill the MHD scaling, with the exception of the Peclet number which for the laboratory case is $\sim$1. 

The parameters for scaled laboratory experiments are in line with plasma conditions achievable on current high-energy density facilities, for instance the OMEGA laser at the U. of Rochester, and future energetic, high-repetition lasers such as ELI-Beamlines in Czech Republic \citep{Jourdain2021ELI}.

\begin{table*}[t]
\caption{MHD scaling of YSO jets and novae to laboratory experiments. The first seven rows are physical (dimensional) parameters, whereas the last six rows are dimensionless parameters. Please refer to the text for further details on the different parameters.}
\begin{tabular}{l|ll|ll}
\hline
Parameter & YSO jet & Scaled experiment & Novae & Scaled experiment \\
 \hline
Length scale $[R]=\rm$ cm & $10^{16}$ & 0.1 & $6\times10^{13}$ & 0.1\\
Density $[n]=\rm  cm^{-3}$ & 10$^{3}$ & $5\times10^{19}$ & 10$^{9}$ & $5\times10^{19}$ \\
Pressure $[P]=\rm  bar$ & 10$^{-13}$ & 10$^{5}$ & $8\times10^{-8}$ & $8\times10^{4}$ \\
Velocity $[v]=\rm km\,s^{-1}$ & 1000 & 700 & 1000 & 1000\\
Magnetic field $[B]=\rm G$ & 10$^{-4}$ & 10$^{5}$ & 10$^{-2}$ & 10$^{4}$\\
Time scale $[t]=\rm s$ & 10$^{8}$ & 10$^{-9}$ & $1.2\times10^{6}$ & $2\times10^{-9}$\\
Temperature $[T]=\rm eV$ & 50 & 1000 & 50 & 1000 \\
\hline
Localization parameter $\delta$ & 10$^{-3}$ & $6\times10^{-1}$ & 10$^{-7}$ & $6\times10^{-1}$\\
Reynolds number $Re$ & 10$^{10}$ & 10$^{4}$ & 10$^{9}$ & 10$^{4}$ \\
Peclet number $Pe$ & 10$^{8}$ & $\sim$1 & 10$^{8}$ & $\sim$1 \\
Magnetic Reynolds number $Re_{M}$ & 10$^{18}$ & 10$^{3}$ & 10$^{17}$ & 10$^{3}$ \\
Euler number $Eu$ & 11 & 8 & 11 & 11 \\
Thermal plasma beta $\beta$ & 50 & 200 & 10$^{4}$ & 10$^{4}$ \\

\label{Tab:1}
\end{tabular}
\end{table*}

\section{Summary and conclusions}\label{conclusions}

We study the interaction regions of YSO jets with an ambient medium and of classical novae outflows with previous ejected material. We show in Section~\ref{jetshocks} that for certain values of the jet/outflow and ambient densities, the working surface in both sources is composed by an adiabatic and a radiative shock. This particular system in which the bow shock is radiative and the reverse shock is adiabatic is the rule in other astrophysical sources such as stellar bow shocks \citep[e.g.,][]{2018ApJ...864...19D}. This shock combination is of interest for particle acceleration and subsequent nonthermal radiation. Particles are expected to be efficiently accelerated in strong adiabatic shocks, while the radiative shock produces a strong compression of the plasma (and magnetic field). 

High-energy particles, accelerated in the reverse, adiabatic shock diffuse up to the dense layer downstream of the radiative shock where can undergo further re energization by compression \citep[e.g.,][]{2011A&A...527A..99E} and also an enhancement of  radiative losses in the denser layer in the form of relativistic Bremsstrahlung for leptons and proton-proton inelastic collisions in the case of hadrons.  We make  order of magnitude estimates for the time scales and gamma-ray luminosity for the case of the YSO jet. In the case of novae, we model the source RS Oph recently detected for the first time in gamma-rays. Our estimations indicate that the reverse shock is adiabatic and might accelerate particles up to high energies that could be responsible for the observed emission.

We found that the parameters for scaled laboratory experiments for YSO jets and nova outflows are in line with plasma conditions achievable in current high-power laser facilities. This opens new laboratory astrophysics working scenarios, especially in the case of novae outflows that was never explored before with this approach.
   
\begin{acknowledgements}
The authors thank the anonymous referee for carefully reading our manuscript and for her/his insightful comments and suggestions. M.~V.~d~V. is supported by the Grants 2019/05757-9 and 2020/08729-3, Fundação de Amparo à Pesquisa do Estado de São Paulo (FAPESP). A.T.A. thanks the Czech Science Foundation under the grant GA\v{C}R 20-19854S 
and the Marie Sk\l{}odowska-Curie fellowship. F.S.V. acknowledges the support from The Royal Society (UK) through a University Research Fellowship.    
\end{acknowledgements}

\bibliographystyle{aa} % style aa.bst
\bibliography{yso.bib} % your references Yourfile.bib

\end{document}